\documentclass[journal]{IEEEtran}
\usepackage{graphicx}
\usepackage{subfig} 

\usepackage{changes}
\usepackage{amsmath}
\usepackage{graphicx}
\usepackage{multicol}
\usepackage{amssymb}
\usepackage{subfig}
\usepackage{lineno}
\usepackage{makecell}
\usepackage{indentfirst}
\usepackage{bm}
\usepackage{color}
\usepackage{cite}
\usepackage{epstopdf}
\usepackage[noend]{algpseudocode}
\usepackage{algorithmicx,algorithm}
\usepackage{amssymb}
\usepackage{xcolor}

\usepackage{mathtools}

\newtheorem{theorem}{Proposition}

\newtheorem{remark}{Remark}
\newtheorem{corollary}{Corollary}

\ifCLASSINFOpdf
  
\else
 
\fi

\usepackage{amsmath}
 
\begin{document}
 
\title{Regularized Covariance Estimation  for  Polarization Radar Detection in Compound Gaussian Sea Clutter}

\author{Lei~Xie,
		Zishu~He,
	   Jun~Tong,
	   Tianle~Liu,
	   Jun~Li and~Jiangtao Xi 
\thanks{This work is supported by the HKUST-BDR Joint Research Institute (HBJRI) Frontier Technology Project under Grant OKT22EG04.}
\thanks{L. Xie is with the Department of Electronic and Computer Engineering, the Hong Kong University of Science and Technology, Hong Kong (e-mail: eelxie@ust.hk). Z. He and J. Li are with the School of Information and Communication Engineering, University of Electronic Science and Technology of China, Chengdu 611731, China. J. Tong and J. Xi are with the School of Electrical, Computer, and Telecommunications
Engineering, University of Wollongong, Wollongong, NSW 2522, Australia. T. Liu is with the School of Communications Engineering, Hangzhou Dianzi University, Hangzhou 310018, China.}
}


\maketitle

\begin{abstract} 

This paper investigates regularized estimation of Kronecker-structured covariance matrices (CM)  for  polarization radar in sea clutter scenarios where the data are assumed to follow the complex, elliptically symmetric (CES) distributions with a Kronecker-structured CM.
To obtain a well-conditioned estimate of the CM, we add penalty terms of Kullback-Leibler divergence to the negative log-likelihood function of the associated complex angular Gaussian (CAG) distribution. This is shown to be equivalent to regularizing Tyler's fixed-point equations by shrinkage. A sufficient condition that the solution exists is discussed. An iterative algorithm is applied to solve the resulting fixed-point iterations and its convergence is proved. 
In order to solve the critical problem of  tuning the shrinkage factors, we then introduce two methods by exploiting oracle approximating shrinkage (OAS) and cross-validation (CV). 
The proposed estimator, referred to as the robust shrinkage Kronecker estimator (RSKE),  is shown to achieve better performance compared with several existing methods when the training samples are limited. 
Simulations are conducted for validating the RSKE 
and demonstrating its high performance by using the IPIX 1998 real sea data. 
\end{abstract}


%
\IEEEpeerreviewmaketitle

\section{Introduction}

\IEEEPARstart{T}{arget} detection in the different scenarios (embracing land, space, atmosphere and seas) is a fundamental problem in radar \cite{9279243,8370234,8752101,7180351,5325631,zhang2019ship,9085903}. However, the presence of clutter poses significant challenges, especially in the sea scenario where the heterogeneity of clutter is particularly significant. Therefore, the sea clutter suppression is a recurrent topic for target detection \cite{7522075,8506468,7126972,1610833}.

Polarization refers to the orientation of the electric and magnetic fields in the plane perpendicular to the direction of wave propagation. Multiple polarization states of a signal can provide more information of a target. 
The resulting polarization diversity has proven to be a useful tool for radar detection in the presence of clutter, especially when discrimination via Doppler frequency is not possible \cite{7180351,9420730,9367301,1661791,7150390,6472767,1292147, XIE2020107401}. 
In polarization array radar, the steering vector can be expressed as the Kronecker product of a polarization component and a space-time component. 

{C}{ovariance} matrix (CM)  estimation is at the core of the target detection \cite{628782, tadjudin1999covariance,4101326,bickel2008regularized,CHEN199910,4104190}. 
The most common CM estimator is the sample covariance matrix (SCM), which is the maximum likelihood estimator (MLE) of the CM for Gaussian data.
However, the Gaussian model does not fit the real sea clutter well due to its heavy tail.
Instead, the compound-Gaussian (CG) distributions, which is a subclass of the complex elliptically symmetric (CES) distributions \cite{6263313}, have been widely used in modeling the sea clutter returns in radar applications \cite{89019,4647137,766928,766939}. 
The SCM suffers poor performance for data with outliers or heavily-tailed distributions due to the lack of robustness.  
To tackle the heavily tailed data, one class of approaches is to censor the training samples with the aim to exclude outliers from the CM estimation \cite{wu2014training, 6825699,
6819451,5570985,1717732,7887259,8401913}. Another class of methods is based on robustification. In particular, for CES distributions, various robust CM estimators based on the M-estimator have been developed and characterized \cite{Huber1964Robust, doi:10.1080/01621459.1974.10482962,10.2307/2241079,10.2307/2957994,4359541,6512056,6636083}. With such estimators, outlying training samples are usually given small weights when an estimate of the CM is produced.

The SCM also requires an abundant number of samples to achieve satisfactory performance. 
Many modern applications involve high-dimensional variables whose statistical characteristics remain stationary over a short observation period, where the large sample support assumption does not hold.
Regularization provides an effective strategy to improve the CM estimation for addressing the challenge of training shortage. 
In particular, a class of linear shrinkage algorithms have been introduced \cite{506799, LEDOIT2004365, 4490277, 5484583} and their integration into robust CM estimators for CES-distributed data have been investigated in the recent works \cite{5743027,6894189,6879466,6913007,xie2021cross}. 
These algorithms estimate the CM by shrinking an estimate of the CM $\widehat{\mathbf{\Sigma}}$ toward a better-conditioned target matrix $\mathbf{T}$. 
There can be various choices for $\widehat{\mathbf{\Sigma}}$ and  $\mathbf{T}$. 
For example, one can choose $\widehat{\mathbf{\Sigma}}$ as the SCM and Tyler's estimator \cite{10.2307/2241079} for Gaussian and non-Gaussian data, respectively. Moreover, different types of target matrices $\mathbf{T}$ can be used, including the identity and diagonal targets. 
The linear shrinkage estimators  can reduce the requirement of samples and provide positive-definite CM estimates. 
The choice of shrinkage factors is a fundamental problem for shrinkage estimators. Various criteria and methods have been studied. In particular, Ledoit and Wolf (LW) propose an approach that asymptotically minimizes the mean squared error (MSE) \cite{LEDOIT2004365}. Then \cite{5484583} improves the LW approach using the Rao-Blackwell theorem and designs the Rao-Blackwell Ledoit and Wolf (RBLW) estimator. The oracle approximating shrinkage (OAS) method is proposed in \cite{5484583}. Both estimators have closed-form expressions and are easily computed. The problem of determining the shrinkage factors can also be cast as a model selection problem and thus generic model selection techniques such as cross-validation (CV) \cite{arlot2010survey} can be applied. The main challenges faced by CV include the choice of the cost function and the heavy computational cost in its direct implementation. {Some efforts are made in \cite{TONG2018223, 7422135} to address these challenges for linear shrinkage estimators with unstructured CM}.

Due to the independence between space-time domain and polarization domain, the polarization-space-time CM also has the Kronecker structure\cite{1310021,7299689,CUI2012430,XIE2020107401}.
Exploiting this structural knowledge about the CM can also significantly reduce the number of unknown parameters and improve its high estimation accuracy under limited training data  \cite{6298979,7439863,8450037,6557488,9054969,8747386,6867011,892662,6558039}.
Particularly, \cite{6298979} proposes a robust estimator for  Kronecker-structured CM and proves that a globally optimal solution can be found, \cite{7439863}  proposes a majorization minimization (MM) solution to the Kronecker maximum likelihood estimator (KMLE), and \cite{LU2005449} introduces the maximum likelihood (ML) estimation of Kronecker-structured CM with the presence of Gaussian clutter. An extension of KMLE is also studied for compound Gaussian clutter with inverse Gamma-distributed texture and Kronecker normalized sample covariance matrix (KNSCM) is proposed in \cite{7950961} to estimate the CM. Although both KMLE and KNSCM provide considerable performance with abundant samples, they still noticeably suffer from performance degradation when the samples are limited.

\subsection{Contributions}
In this paper, we consider the estimation of Kronecker-structured CM for polarized sea clutter data under low sample supports.  
In order to improve the performance in this case, we introduce the Kullback-Leibler divergence penalty to the negative log-likelihood function for the CM estimation. We then  derive a robust shrinkage Kronecker estimator (RSKE) that aims to achieve well-conditioned\footnote{For a positive-definite, Hermitian matrix, the condition number is defined as the ratio of its maximum and minimum eigenvalues \cite{823928}. A well-conditioned matrix indicates that its condition number is small.} and highly accurate CM estimates. With RSKE, the structural knowledge is exploited together with robustification and regularization techniques. Based on the findings of the previous studies in \cite{6263313, 7547360, 6879466, 5743027, wiesel2015structured, TONG2018223} and others, we investigate the existence of RSKE, its iterative solver and convergence, and also the choice of the shrinkage factors. We then study the performance of the RSKE for the polarization-space-time adaptive processing (PSTAP) in radar applications. The contributions of this paper can be summarized as follows:
\begin{enumerate}
\item   We propose to apply robust shrinkage Kronecker estimator (RSKE) to polarization radar detection in compound Gaussian sea clutter. We  show that the RSKE can be interpreted as the minimizer of a negative log-likelihood function penalized by the Kullback-Leibler divergence. Based on this, the condition for the existence of RSKE is established under some mild assumptions, which provides insights to the relationship between the dimensionality, sample size and shrinkage factors. 
\item We study an iterative solver involving two fixed-point equations to find RSKE and prove its convergence. Following the majorization-minimization framework, we prove the monotonic decrease of the penalized log-likelihood function over iterations. We show that, with fixed shrinkage factors and arbitrary positive-definite initial estimates, the iterative solver converges. 
\item     
We address the critical challenge of shrinkage factor choice in order to exploit the potential of RSKE. We introduce data-driven methods that automatically tune the linear shrinkage factors, based on oracle approximating shrinkage (OAS) and cross-validation (CV).  
The OAS method adopts a minimum MSE (MMSE) criterion and plug-in estimates of the oracle shrinkage factors. For the CV methods, we start with a quadratic loss  
for leave-one-out CV (LOOCV) and derive analytical solutions of the shrinkage factors  
which can approach the performance of the oracle solutions that minimize the MSE  of CM estimation. The complexities of these different methods are analyzed. 
It is found that the analytical CV solutions successfully address the key challenge of high computational complexity of general applications of CV, and the resulting RSKE has a complexity similar to that of the KMLE. 
\end{enumerate}

\subsection{Organization}
The remainder of this paper is organized as follows. 
Section II introduces the signal model, the RSKE as well as its existence and iterative solution. 
Section III gives the choices of the shrinkage factors.
Section IV presents simulation results to show the performance of CM estimation. 
Finally, Section V gives the conclusions.

\section{Robust Shrinkage Kronecker Estimator (RSKE)}

In this section, we introduce the robust shrinkage estimator for Kronecker-structured covariance matrices. We first discuss the motivation, then give the condition for its existence, and finally introduce the iterative solver and its convergence property.

\subsection{Signal Model}
Consider a pulsed Doppler radar deploying a uniform linear array (ULA) of $N_s$ antennas, each of which can measure electromagnetic wave in $N_p$ polarization channels \cite{4454222,1310021,7950961}. A burst of $N_t$ identical pulses at a constant pulse repetition frequency (PRF) of $f_r$ are transmitted during the coherent processing interval (CPI). The received signals of all the polarization channels at each sensor in the cell under test (CUT) are down-converted to baseband or to an intermediate frequency in all the pulses at each sensor. They are then processed by the corresponding matched filters and sampled and stacked into an $N$-dimensional vector $\mathbf{y} \in \mathbb{C}^{N\times 1}$, where $N=N_p N_t N_s$. Let $\mathbf{y}_l, l=1,2,\cdots, L$, be $L$ independent, identically distributed (i.i.d) signal-free secondary data, arising from adjacent range cells. 

 Radar detection is a binary hypothesis testing problem, where hypotheses $H_0$ and $H_1$ correspond to target absence and presence, respectively. We first ignore noise in the received signal, which approximates the case of high clutter-to-noise-ratio (CNR). The received signal can then be approximately modeled as \cite{1561887,7950961} 
\begin{equation} \label{pstapmodel}
	\left\{
	\begin{array}{ll}
	H_0: &\mathbf{y}=\mathbf{c}_0\\
	&\mathbf{y}_l=\mathbf{c}_l, l=1,2,\cdots,L\\
	H_1: &\mathbf{y}=\alpha\mathbf{s}+\mathbf{c}_0\\
	&\mathbf{y}_l=\mathbf{c}_l, l=1,2,\cdots,L\\
	\end{array}
	\right.
\end{equation} 
where $\alpha$ denotes the complex amplitude of the target signal, $\mathbf{s}$ denotes the steering vector of target, and $\{\mathbf{c}_l\}$ denote the clutter returns in the CUT and adjacent cells. In sea clutter scenarios, experimental trials have shown a good fitting of the compound Gaussian model to the heterogeneous clutter measurements \cite{4647137,766939}. The received clutter can then be modeled using a positive texture and a Gaussian vector  referred to as the speckle, i.e.,
\begin{equation}
	\left\{
	\begin{array}{ll}
	 \mathbf{c}_0= \sqrt{\tau_0} \mathbf{u}_0 \in \mathbb{C}^{N\times 1},  \\	
	  \mathbf{c}_l = \sqrt{\tau_l} \mathbf{u}_l \in \mathbb{C}^{N\times 1},   
	\end{array}
	\right.
\end{equation} 
where $\tau_l $ is the texture and  $\mathbf{u}_l$ the speckle component.  
We assume $\mathbb{E} (\tau_l) < \infty, \forall l, $ so that the CM of $\mathbf c_l$ exists, where $\mathbb{E}(\cdot)$ denotes the mathematical expectation. 
We assume that all the clutter patches are associated with the same terrain and thus $\mathbf{u}_l$ are zero-mean and  i.i.d with a shared covariance matrix $\mathbf{R}$, i.e., $\mathbf{u}_l \sim \mathcal{CN}(\mathbf{0}, \mathbf{R})$.
For conciseness, we here drop the subscript $l$ of $\mathbf u_l$ while discussing its covariance matrix $\mathbf R$ below.

The clutter signal for a polarimetric radar can be expressed as the sum of $N_c$ clutter patches in the same range cell, i.e.,
\begin{equation}
    \begin{split}
	&\mathbf{u}=\sum_{i=1}^{N_c} \mathbf{a}_{c}^{(i)} \otimes \mathbf{p}_{c}^{(i)} \in \mathbb{C}^{N\times 1},
    \end{split}
   \label{xl}\end{equation}
where $\mathbf{a}_{c}^{(i)} \in \mathbb{C}^{N_sN_t \times 1}$ and  $\mathbf{p}_{c}^{(i)} \in \mathbb{C}^{N_p \times 1}$ denote the space-time steering vector and polarization scattering vector of the $i$th clutter patch, respectively. 
Similarly, the polarization-space-time steering vector of the target can be written as $\mathbf s =\mathbf a_{t} \otimes \mathbf{p}_{t}$, where $\mathbf a_t $ denotes the target space-time steering vector which depends on the direction and velocity of the target.  

Following \cite{249129}, we assume that 
$N_p=3$ and $\mathbf{p}_{c}^{(i)}$ consists of three complex elements: HH, VV, and HV, i.e., 
\begin{equation}
    \begin{split}
	\mathbf{p}_{c}^{(i)}=\left[ p_{c,\textit{hh}}^{(i)}, p_{c,\textit{vv}}^{(i)},p_{c,\textit{hv}}^{(i)} \right]^\mathrm{T},
    \end{split}
\end{equation}
where $\left( \cdot \right)^\mathrm{T}$ denotes the transpose. 
Furthermore, we assume $\mathbf{p}_{c}^{(i)}$ follows a complex Gaussian distribution with zero mean and covariance matrix \cite{249129,7950961} 
\begin{equation}
    \begin{split}
  \mathbb{E}\left(\mathbf{p}_{c}^{(i)}\left(\mathbf{p}_{c}^{(i)} \right)^\mathrm{H} \right)=\varepsilon_i \left[\begin{matrix}
   1 & \rho_{c} \sqrt{\gamma_{c}} & 0\\
\left(\rho_{c}\right)^{*}\sqrt{\gamma_{c}} & {\gamma_{c}} & 0\\
0 & 0 & \delta_{c}\\
  \end{matrix}\right],
    \end{split}
\end{equation}
with  $\left( \cdot \right)^\mathrm{H}$ denoting the conjugate transpose, 
$\varepsilon_i=\mathbb{E}\left( \left\vert p_{c,\textit{hh}}^{(i)} \right\vert^2 \right)$, $\delta_{c}=\frac{\mathbb{E}\left( \left\vert p_{c,\textit{hv}}^{(i)} \right\vert^2 \right)}{\mathbb{E}\left( \left\vert p_{c,\textit{hh}}^{(i)} \right\vert^2 \right)}$, $\gamma_{c}=\frac{\mathbb{E}\left( \left\vert p_{c,\textit{vv}}^{(i)} \right\vert^2 \right)}{\mathbb{E}\left( \left\vert p_{c,\textit{hh}}^{(i)} \right\vert^2 \right)}$ and $\rho_{c}=\frac{\mathbb{E}\left(  p_{c,\textit{hh}}^{(i)} \left( p_{c,\textit{vv}}^{(i)} \right)^{*} \right)}{\left[ \mathbb{E}\left( \left\vert p_{c,\textit{hh}}^{(i)} \right\vert^2 \right) \mathbb{E}\left( \left\vert p_{c,\textit{vv}}^{(i)} \right\vert^2 \right)\right]^{1/2}}$.

The space-time steering vector is expressed as
\begin{equation}
    \begin{split}
  \mathbf{a}_{c}^{(i)}=\mathbf{a}_d \left( f_{d,i} \right) \otimes  \mathbf{a}_s \left( f_{s,i} \right)\in \mathbb{C}^{N_sN_t \times 1}, 
   \label{afdfs}
    \end{split}
\end{equation}
where $f_{d,i}=\left( 2v_a/\lambda f_r \right)\cos(\phi_i) $ denotes the normalized Doppler frequency, $f_{s,i}=\left( d/\lambda \right)\cos(\phi_i) $ the normalized spatial frequency, $d$ the inter-element spacing, $v_a$ the velocity of the platform, $\lambda$ the radar wavelength, $\phi_i$ the direction of $i$th clutter patch with respect to the array, $\otimes$ the Kronecker product, and 
\begin{equation}
    \begin{split}
   &\mathbf{a}_d \left( f_{d,i} \right)=\left[ 1,e^{j2\pi f_{d,i}},\cdots,e^{j2\pi (N_t-1) f_{d,i}} \right]^{\mathrm{T}}\in \mathbb{C}^{N_t \times 1},\\
   &\mathbf{a}_s \left( f_{s,i} \right)=\left[ 1,e^{j2\pi f_{s,i}},\cdots,e^{j2\pi (N_s-1) f_{s,i}} \right]^{\mathrm{T}}\in \mathbb{C}^{N_s \times 1},
   \label{adas}
    \end{split}
\end{equation}
are the temporal and spatial steering vectors, respectively.

The covariance matrix of $\mathbf{u}$   can be given as \cite{7950961}
\begin{equation}
    \begin{split}
    \mathbf{R}=\mathbb{E} \left( \mathbf{u}  \mathbf{u} ^{\mathrm{H}} \right)= \mathbf{R}_\textit{st} \otimes \mathbf{R}_p  \in \mathbb{C}^{N \times N}, 
    \end{split}
    \label{Rxpst}
\end{equation}
where  the space-time and  polarization covariance matrices are respectively defined as
\begin{equation}
    \begin{split}
\mathbf{R}_\textit{st} \triangleq \sum_{i=1}^{N_c} \varepsilon_i  \mathbf{\alpha}_{c}^{(i)} \left( \mathbf{\alpha}_{c}^{(i)} \right)^\mathrm{H}\in \mathbb{C}^{N_sN_t \times N_sN_t}, 
    \end{split}
\label{Rst}
\end{equation}
and
\begin{equation}
    \begin{split}
\mathbf{R}_p \triangleq \left[\begin{matrix}
   1 & \rho_{c} \sqrt{\gamma_{c}} & 0\\
\left(\rho_{c}\right)^{*}\sqrt{\gamma_{c}} & {\gamma_{c}} & 0\\
0 & 0 & \delta_{c}\\
	\end{matrix}\right]\in \mathbb{C}^{3 \times 3}. 
    \end{split}
\label{Rp}
\end{equation}

\subsection{Kronecker Maximum Likelihood Estimator}
 The CES distributions have been widely employed for modeling radar clutter and many previous experiments have shown that they fit the measured clutter well \cite{89019,4647137,766928,766939,6263313,6636083}. Therefore, following these studies and as will also be demonstrated in Section IV,
  we  assume that the sea clutter $\mathbf{y}_l$ follows the CES distribution. 
 The  probability density function (p.d.f.) of $\mathbf{y}_l$ is of the form
\begin{equation}
    \begin{split}
	p(\mathbf{y}_l)=C_{N,g} \det(\mathbf{R})^{-1} g\left(  \mathbf{y}_l^\mathrm{H}   \mathbf{R}^{-1} \mathbf{y}_l \right), 
    \end{split}
\label{PDFmodel1}
\end{equation}
where $g(\cdot)$ denotes the density generator and $C_{N,g}$ a normalizing constant. Note that $\mathbf{R}$ is also known as the scatter matrix \cite{6263313,6636083}.  

The normalized samples $\{\mathbf{x}_l=\frac{\mathbf{y}_l}{\Vert \mathbf{y}_l \Vert}\}_{l=1}^L$, which belong to a complex unit $N$-dimensional sphere, follows the complex angular Gaussian (CAG) distribution \cite{6636083,6263313}. 
The joint distribution function of $\{\mathbf{x}_l\}_{l=1}^L$  is expressed as \cite{6263313}
\begin{equation}
    \begin{split}
	p(\{\mathbf{x}_l \})=\prod_{l=1}^L p(\mathbf{x}_l)\propto \mathrm{det}(\mathbf{R})^{-L} \prod_{l=1}^L \left(  \mathbf{x}_l^\mathrm{H} \mathbf{R}^{-1}  \mathbf{x}_l\right)^{-N},
   \end{split}
\end{equation}
where $\mathrm{det}(\cdot)$ denotes the  determinant.  
After omitting some additive constants and scaling, the negative log-likelihood function of such a joint distribution is given by
\begin{equation}
    \begin{split}
\mathcal{L}_0 \left(  \widehat{\mathbf{R}}_{\mathit{st}}, \widehat{\mathbf{R}}_p  \right)&= \log \det \left(  \widehat{\mathbf{R}}_{\mathit{st}}  \otimes \widehat{\mathbf{R}}_p  \right) \\
&+ \frac{N}{L}\sum_{l=1}^{L} \log  \mathbf{y}_l^\mathrm{H} \left(\widehat{\mathbf{R}}_{\mathit{st}}\otimes \widehat{\mathbf{R}}_p \right)^{-1}\mathbf{y}_l , 
    \end{split}
    \label{ProblemMLE}
\end{equation}
where $\widehat{\mathbf{R}}_{\mathit{st}} \in \mathbb{S}_{++}^{N_{\mathit{st}}}$, $\widehat{\mathbf{R}}_{p} \in \mathbb{S}_{++}^{N_p}$, and we have used the fact that  
$
\log  ( \mathbf{y}_l^\mathrm{H} \widehat{\mathbf{R}}^{-1}\mathbf{y}_l  ) - \log  ( \mathbf{x}_l^\mathrm{H} \widehat{\mathbf{R}}^{-1}\mathbf{x}_l  ) = \log(||\mathbf y_l||^2)
$ is irrelevant to $\widehat{\mathbf R}=  \widehat{\mathbf{R}}_{\mathit{st}}  \otimes \widehat{\mathbf{R}}_p$ in the likelihood function.
 {The above cost function $\mathcal{L}_0  (  \widehat{\mathbf{R}}_{\mathit{st}}, \widehat{\mathbf{R}}_p   )$ is non-convex in the classical definitions but is jointly
g-convex (geodesic-convex) \cite{6298979} with respect to $\widehat{\mathbf{R}}_{\mathit{st}}$ and $\widehat{\mathbf{R}}_p$. 
Minimizing this cost function produces the KMLE \cite{wiesel2015structured,7439863}.
In the low-sample-support cases, the solution of KMLE can suffer from significant errors and ill-conditioning. 
For many applications such as beamforming and spectral estimation \cite{622504, habets2009new, 479429,book2006,1263229,9052470,9444455}, the inverse of the CM estimate is required. 
Inverting an erroneous, ill-conditioned CM estimate can bring enormous errors. This motivates the design of accurate, well-conditioned CM estimators. 

\subsection{Regularization via KL Divergence Penalty} 
In this subsection, we introduce a penalized estimator that promotes well-conditioned estimates of the sub-CMs $\mathbf R_{\mathit{st}}$ and $\mathbf R_p$. 
We adopt penalty terms of the Kullback-Leibler divergence {for Gaussian distributions} \cite{davis2007information}, i.e., 
\[ D_{\mathrm{KL}}\left( \mathbf{X},\mathbf{Y}\right)=\mathrm{Tr}\left( \mathbf{X}\mathbf{Y}^{-1} \right)-\log \det \left( \mathbf{X}\mathbf{Y}^{-1} \right)-N,\]
where $\mathbf{X},\mathbf{Y} \in \mathbb{S}_{++}^{N}$. 
As shown in  \cite{dhillon2008log}, the KL divergence  
$D_{\mathrm{KL}}\left( \mathbf{X},\mathbf{I}_N\right)$ can effectively constrain the condition number of $\mathbf X$.
We thus add the penalty terms $\alpha_{\mathit{st}} D_{\mathrm{KL}} ( \widehat{\mathbf{R}}_{\mathit{st}}^{-1},\mathbf{I}_{N_{\mathit{st}}})$ and $\alpha_p D_{\mathrm{KL}} ( \widehat{\mathbf{R}}_p^{-1},\mathbf{I}_{N_p})$ to the negative log-likelihood function in (\ref{ProblemMLE}) to promote well-conditioned estimates $\widehat{\mathbf{R}}_{\mathit{st}}$ and $\widehat{\mathbf{R}}_p$, where $\alpha_{\mathit{st}}=\frac{N_p \rho_{\mathit{st}}}{1-\rho_{\mathit{st}}}$ and $\alpha_p=\frac{N_{\mathit{st}} \rho_p}{1-\rho_p}$ with $\rho_{\mathit{st}} \in [0,1)$ and $\rho_p \in [0,1)$.
Ignoring some additive constants which are irrelevant to $\widehat{\mathbf{R}}_{\mathit{st}}$ and $\widehat{\mathbf{R}}_p$, the penalized negative log-likelihood function is obtained as
 \begin{equation}
    \begin{split}
&\mathcal{L} \left(  \widehat{\mathbf{R}}_{\mathit{st}}, \widehat{\mathbf{R}}_p  \right) 
=  \frac{N_p}{1-\rho_{\mathit{st}}} \log \det (  \widehat{\mathbf{R}}_{\mathit{st}}  ) +\frac{ N_{\mathit{st}}}{1-\rho_p} \log \det (  \widehat{\mathbf{R}}_p  )\\
& + \frac{N}{L} \sum_{l=1}^{L} \log \mathbf y_l^\mathrm{H} \left(\widehat{\mathbf{R}}_{\mathit{st}}\otimes \widehat{\mathbf{R}}_p \right)^{-1} \mathbf y_l + \frac{N_p \rho_{\mathit{st}}}{1-\rho_{\mathit{st}}} \mathrm{Tr} \left(  \widehat{\mathbf{R}}_{\mathit{st}}^{-1}  \right)\\
& + \frac{N_{\mathit{st}} \rho_p}{1-\rho_p}  \mathrm{Tr} \left(  \widehat{\mathbf{R}}_p^{-1}  \right),  
    \end{split}
    \label{KMLEProblem0}
\end{equation} 
  which reduces to $\mathcal{L}_0  (  \widehat{\mathbf{R}}_{\mathit{st}}, \widehat{\mathbf{R}}_p   )$ in (\ref{ProblemMLE}) when $\rho_{\mathit{st}}=\rho_p=0$.
By adding the penalty terms which are convex, the obtained objective function is also g-convex w.r.t. $\widehat{\mathbf{R}}_{\mathit{st}}$ and $\widehat{\mathbf{R}}_p$. This guarantees that all local minimizers of $\mathcal{L}  (  \widehat{\mathbf{R}}_{\mathit{st}}, \widehat{\mathbf{R}}_p   )$ are also globally optimal, following \cite[Proposition 1]{6298979}.
Minimizing the penalized log-likelihood function by setting 
${ \partial \mathcal{L}  (  \widehat{\mathbf{R}}_{\mathit{st}}, \widehat{\mathbf{R}}_p   )}/{\partial \widehat{\mathbf{R}}_{\mathit{st}}} = \mathbf 0$
and ${\partial \mathcal{L}  (  \widehat{\mathbf{R}}_{\mathit{st}}, \widehat{\mathbf{R}}_p   )}/{\partial \widehat{\mathbf{R}}_p}= \mathbf 0$ 
yields the fixed-point equations
\begin{subequations}
\label{fixpoint2}
\begin{equation}
    \begin{split}
\widehat{\mathbf{R}}_{\mathit{st}}=(1-\rho_{\mathit{st}}) \frac{N_{\mathit{st}}}{ L}\sum_{l=1}^L \frac{\mathbf{Y}_l^\mathrm{H} \widehat{\mathbf{R}}_{p}^{-1} \mathbf{Y}_l}{\mathbf{y}_l^\mathrm{H} \left( \widehat{\mathbf{R}}_{\mathit{st}}^{-1} \otimes \widehat{\mathbf{R}}_{p}^{-1} \right)  \mathbf{y}_l}+\rho_{\mathit{st}}\mathbf{I}_{N_{\mathit{st}}},
\end{split}
\label{fixpoint21}
\end{equation} 
\begin{equation}
    \begin{split}
\widehat{\mathbf{R}}_{p}=(1-\rho_p) \frac{N_p}{ L} \sum_{l=1}^L \frac{\mathbf{Y}_l \widehat{\mathbf{R}}_{\mathit{st}}^{-1} \mathbf{Y}_l^\mathrm{H}}{\mathbf{y}_l^\mathrm{H} \left( \widehat{\mathbf{R}}_{\mathit{st}}^{-1} \otimes \widehat{\mathbf{R}}_{p}^{-1} \right)  \mathbf{y}_l}+\rho_p\mathbf{I}_{N_p}. 
\end{split}
\label{fixpoint22}
\end{equation} 
\end{subequations}
In the above, we have defined 
\begin{equation}
    \begin{split}
&\mathbf{Y}_l = \mathrm{unvec}_{N_p N_{\mathit{st}}}(\mathbf{y}_l) \\  & 
\triangleq \left[\begin{matrix}
{y}_l^{(1)} & {y}_l^{(N_p+1)} & \cdots & {y}_l^{(N_p(N_{\mathit{st}}-1)+1)}\\
{y}_l^{(2)} & {y}_l^{(N_p+2)} & \cdots & {y}_l^{(N_p(N_{\mathit{st}}-1)+2)}\\
\vdots & \vdots & \ddots & \vdots\\
{y}_l^{(N_p)} & {y}_l^{(2 N_p)} & \cdots & {y}_l^{(N_p(N_{\mathit{st}}-1)+N_p)}\\
\end{matrix}\right]\in \mathbb{C}^{N_p \times N_{\mathit{st}}},
    \end{split}
\label{Yl}
\end{equation} 
where ${y}_l^{(i)}$ denotes the $i$th entry of $\mathbf{y}_l$ and $\mathrm{unvec}_{N_p N_{\mathit{st}}} (\cdot)$ reshapes a vector into a $N_p \times N_{\mathit{st}}$ matrix as shown above. 
Therefore, the solution to (\ref{fixpoint2}), if exists, can be interpreted as the minimizer of the penalized negative log-likelihood function (\ref{KMLEProblem0}). These fixed-point equations interestingly have the same form as the linear shrinkage estimators for unstructured CM \cite{LEDOIT2004365, 5484583, 5743027,6894189,6879466,6913007}. 
Following these work, we refer to the resultant CM estimator as the robust shrinkage Kronecker estimator (RSKE), with shrinkage factors $\rho_{\mathit{st}}$ and $\rho_p$. 
The KMLE \cite{7439863} can be obtained as a special case of RSKE by letting $\rho_{\mathit{st}}= \rho_p=0$.  

It should be noted that in \cite{wiesel2015structured},  estimators that exploit robustification and shrinkage for the unstructured CM and robust estimators for the Kronecker-structured CM have been studied via the geodesic convexity. 
The KL divergence penalty has also been exploited in \cite{6879466} for robust estimation of unstructured CM. We here extend these studies to the estimation of Kronecker-structured CM  by simultaneously exploiting robustification and shrinkage.

\subsection{Existence of RSKE}

In this subsection, we examine the conditions under which the RSKE exists. When $\rho_{\mathit{st}}$ and $\rho_p$ are small, it is possible that the cost function (\ref{KMLEProblem0}) tends to  $-\infty$ on the boundary of the set $\mathbb{S}_{++}^{N_{\mathit{st}}}$ and  $\mathbb{S}_{++}^{N_p}$, i.e., $(\ref{KMLEProblem0})$ becomes unbounded below and there is no solution to the fix-point equations of (\ref{fixpoint2}).
The existence of the shrinkage Tyler's estimator for unstructured CM has been studied in \cite{6879466}, where the relationship between the shrinkage factors, sample size, and dimensionality is revealed. 
By establishing the condition under which the cost function tends to $+\infty$ on the boundary of the set of positive-definite, Hermitian matrix, the minimum shrinkage factor for the existence of the CM estimator is obtained \cite{6879466}. 
This result, however, can not directly determine the conditions of the two shrinkage factors affecting each other. 
In this work, we follow \cite[Theorem 3]{6879466} and its proof to study the RSKE. 
We first construct auxiliary functions by which the penalized negative log-likelihood function (\ref{KMLEProblem0}) can be lowerbounded. The two auxiliary functions have a similar form as (15) in \cite{6879466}. Thus, using the same treatment of \cite{6879466}, we can examine the conditions for the auxiliary functions tending to $+\infty$ at the boundary. Based on the results, we 
can obtain the following sufficient condition for the existence of a solution to the RSKE:

\begin{theorem}
\label{ExistenceMain}
The cost function (\ref{KMLEProblem0}) {has a {finite} lower bound over the set of positive-definite $\widehat{\mathbf{R}}_{\mathit{st}}$ and $\widehat{\mathbf{R}}_{p}$}, i.e., a solution to (\ref{fixpoint2}) exists if the following conditions are satisfied:
\begin{enumerate}
\item[(1)] None of $\mathbf{r}_{j,l}$ and $\mathbf{c}_{i,l}$ is an all-zero vector, where $\mathbf{r}_{j,l}\in \mathbb{C}^{N_{\mathit{st}}\times 1}$ denotes the 
$j$th row of $\mathbf{Y}_l$ and $\mathbf{c}_{i,l}\in \mathbb{C}^{N_p\times 1}$ denotes the $i$th column of $\mathbf{Y}_l$;
\item[(2)] There exist $\beta_1 \in[0,1],\beta_2 \in[0,1]$ with $\beta_1+\beta_2=1$ such that for any proper subspace $\mathcal{S}_{\mathit{st}}\subset\mathbb{C}^{N_{\mathit{st}}\times 1}$ and $\mathcal{S}_p \subset\mathbb{C}^{N_p\times 1}$ in the space of length-$N_{\mathit{st}}$ and  -$N_p$ vectors, respectively, 
 \begin{subequations}
\label{conditionA21}
 \begin{equation}
    \begin{split}
    P_{L N_p}\left(\mathcal{S}_{\mathit{st}} \right) < \frac{ \left(L N_p+\alpha_{\mathit{st}} L \right) \mathrm{dim} (\mathcal{S}_{\mathit{st}})-\beta_2 L N}{\beta_1   L N },
     \end{split}
     \label{conditionAA21}
\end{equation}
 \begin{equation}
    \begin{split}
    P_{L N_{\mathit{st}}}\left(\mathcal{S}_p \right) < \frac{ \left(L N_{\mathit{st}}+\alpha_p L \right) \mathrm{dim} (\mathcal{S}_p)-\beta_1 L N}{\beta_2   L N },
     \end{split}
     \label{conditionAB21}
\end{equation}
\end{subequations}
where $P_{L N_p}(\mathcal{S}_{\mathit{st}})\triangleq \frac{\sum_{j=1}^{N_p} \sum_{l=1}^{L}  1_{\mathbf{r}_{j,l}\in \mathcal{S}_{\mathit{st}}}}{L N_p},$ $P_{L N_{\mathit{st}}}(\mathcal{S}_p)\triangleq \frac{\sum_{i=1}^{N_{\mathit{st}}} \sum_{l=1}^{L}  1_{\mathbf{c}_{i,l}\in \mathcal{S}_p}}{L N_{\mathit{st}}},$
$1_x$ denotes the indicator function. 
\end{enumerate}
\end{theorem}

\textit{Proof}: See Appendix \ref{Existence}.

In general, the above conditions require that the number of samples to be sufficiently large, and the samples are  {evenly} spread out in the whole space.
\begin{corollary}
 If the samples are evenly spread out in the whole space, such that $P_{L N_p}(\mathcal{S}_{\mathit{st}}) \leq\frac{ \mathrm{dim}(\mathcal{S}_{\mathit{st}}) }{\min (N_{\mathit{st}}, L N_p)}= \frac{\mathrm{dim}(\mathcal{S}_{\mathit{st}})  \max (N_{\mathit{st}}, L N_p)}{L N}$ and $P_{L N_{\mathit{st}}}(\mathcal{S}_p) \leq \frac{\mathrm{dim}(\mathcal{S}_p) \max (N_p, L N_{\mathit{st}})}{L N}$, then 
 Condition (2) in \emph{Proposition 1} is equivalent to 
\begin{subequations}
   \begin{equation}
    \begin{split}
    \rho_{\mathit{st}}>1-\frac{L N_p}{ \beta_1 \max (N_{\mathit{st}}, L N_p)+ \beta_2 L N},
     \end{split}
     \label{conditionAA4}
\end{equation}
  \begin{equation}
    \begin{split}
    \rho_p>1-\frac{L N_{\mathit{st}}}{ \beta_2 \max (N_p, L N_{\mathit{st}})+ \beta_1 L N}.
     \end{split}
     \label{conditionAB4}
\end{equation}
\end{subequations}
\end{corollary}

\emph{Proof}: Let $\mathrm{dim}(\mathcal{S}_{\mathit{st}}) \triangleq d_{\mathit{st}}$. Recall that $\alpha_{\mathit{st}}=\frac{N_p \rho_{\mathit{st}}}{1-\rho_{\mathit{st}}}$ and $\alpha_p=\frac{N_{\mathit{st}} \rho_p}{1-\rho_p}$. The condition (\ref{conditionAA21}) is satisfied when
  \begin{equation}
    \begin{split}
     \frac{d_{\mathit{st}} \max (N_{\mathit{st}}, L N_p)}{L N} < \frac{ \frac{L N_p d_{\mathit{st}}}{1-\rho_{\mathit{st}}}-\beta_2 L N}{\beta_1   L N }.
     \end{split}
     \label{conditionAA2}
\end{equation}
Rearranging (\ref{conditionAA2}), one has $\rho_{\mathit{st}}>1-\frac{L N_p}{ \beta_1 \max (N_{\mathit{st}}, L N_p)+ \frac{\beta_2 L N}{d_{\mathit{st}}}}$ for arbitrary $d_{\mathit{st}}=1,\cdots, N_{\mathit{st}}-1$, i.e.,
  \begin{equation}
  \nonumber
    \begin{split}
 \rho_{\mathit{st}}&>\max_{d_{\mathit{st}}} \left(1-\frac{L N_p}{ \beta_1 \max (N_{\mathit{st}}, L N_p)+ \frac{\beta_2 L N}{d_{\mathit{st}}}}\right)\\
 &=1-\frac{L N_p}{ \beta_1 \max (N_{\mathit{st}}, L N_p)+ \beta_2 L N},
 \end{split}
\end{equation}
which is exactly (\ref{conditionAA4}). Similarly, we have (\ref{conditionAB4}).

\begin{remark}
 Condition (2) in \emph{Corollary 1} shows the relationship between the shrinkage factors, the number of samples $L$, and the dimension of the sub-CMs $N_{\mathit{st}}$ and $N_p$.  
In general, a larger shrinkage factor  {$\rho_{\mathit{st}}$ is required} when $L$ decreases or $N_{\mathit{st}}$ increases. 
Moreover, Condition (2) can be easily checked. For example, when $\beta_1=1$ and $\beta_2=0$, $ {\rho_{\mathit{st}}} > \max(1-\frac{ L N_p}{  \max(N_{\mathit{st}},L N_p)  },0)$ and $ {\rho_p}  > \max(1-\frac{1}{N_p},0)$.
When $N_p=1$, $N=N_{\mathit{st}}$, the Kronecker-structured CM reduces to an unstructured one. Then Condition (2) becomes $\rho_{\mathit{st}}>1-\frac{L}{\max(N,L)}$ and $\rho_p>0$. When $L\geq N$, the condition is $\rho_{\mathit{st}} \in (0,1)$. When $L<N$, the condition is $\rho_{\mathit{st}} \in (1-\frac{L}{N},1)$, which
agrees with the result in \cite{6894189,6879466} for the case of unstructured CM.  
\end{remark}

 \subsection{Iterative Solver and Its Convergence} 
\label{SecConv}
Similarly to \cite{5743027,6894189,6879466,6913007}, we solve (\ref{fixpoint2})  by applying the process below, which involves two fixed-point iterations: 
\begin{subequations}
\label{SKEiter}
\begin{equation}
    \begin{split}
\widehat{\mathbf{R}}_{\mathit{st}}^{(k+1)} (\rho_{\mathit{st}})&=(1-\rho_{\mathit{st}})\widehat{\mathbf{C}}_{\mathit{st}}^{(k+1)}+\rho_{\mathit{st}}\mathbf{I}_{N_{\mathit{st}}},
\end{split}
\label{iterA}
\end{equation} 
\begin{equation}
    \begin{split}
\widehat{\mathbf{R}}_p^{(k+1)} (\rho_p)&=(1-\rho_p)\widehat{\mathbf{C}}_p^{(k+1)}+\rho_p\mathbf{I}_{N_p},
\end{split}
\label{iterB}
\end{equation} 
\end{subequations}
where
\begin{subequations}
\label{Ckone}
\begin{equation}
    \begin{split}
\widehat{\mathbf{C}}_{\mathit{st}}^{(k+1)}=\frac{N_{\mathit{st}}}{ L} \sum_{l=1}^L \frac{\mathbf{Y}_l^\mathrm{H} {\left(\widehat{\mathbf{R}}_p^{(k)}\right)}^{-1} \mathbf{Y}_l}{\mathbf{y}_l^\mathrm{H} \left( \widehat{\mathbf{R}}_{\mathit{st}}^{(k)} \otimes \widehat{\mathbf{R}}_p^{(k)} \right)^{-1}  \mathbf{y}_l},
\end{split}
\label{CAk1}
\end{equation} 
\begin{equation}
    \begin{split}
\widehat{\mathbf{C}}_p^{(k+1)}=\frac{N_p}{ L}  \sum_{l=1}^L 
\frac{\mathbf{Y}_l \left(\widehat{\mathbf{R}}_{\mathit{st}}^{(k)}\right)^{(-1)} \mathbf{Y}_l^\mathrm{H}}{\mathbf{y}_l^\mathrm{H} \left( \widehat{\mathbf{R}}_{\mathit{st}}^{(k)} \otimes \widehat{\mathbf{R}}_p^{(k)} \right)^{-1}  \mathbf{y}_l}, 
\end{split}
\label{CBk1}
\end{equation} 
\end{subequations}
and $\widehat{\mathbf{R}}_{\mathit{st}}^{(k)}$ and $\widehat{\mathbf{R}}_{p}^{(k)}$ denote the estimates of the sub-CMs at the $k$th iteration. 
 In this paper, we choose the initial CM 
estimates as $\widehat{\mathbf{R}}_{\mathit{st}}^{(0)} = \mathbf I_{N_{\mathit{st}}}$ and $\widehat{\mathbf{R}}_{p}^{(0)}= \mathbf I_{N_p}$ 
for simplicity.

It is useful to examine the convergence property of the above iterative estimator which generalizes Tyler's estimator \cite{10.2307/2241079} and its shrinkage extension \cite{5743027,6879466,6913007} to the case of Kronecker-structured CM. 
The works \cite{10.2307/2241079,5743027,6879466,6913007} assume unstructured CM and thus their solutions can be characterized by a single fixed-point equation. The convergence of the iterative process for Tyler's estimator is proved in \cite{10.2307/2241079} by examining the fixed-point iterations. For the shrinkage  {extension} of Tyler's estimator,  the convergence is proved in \cite{5743027} by applying the concave Perron-Frobenius theory, in \cite{6879466} by applying the majorization-minimization theorem, and in \cite{6913007} by applying the monotone bounded convergence theorem. For the Kronecker-structured CM, though the case of the KMLE has been studied in \cite{6298979}, in this work we incorporate shrinkage into the estimator and the convergence has not been analyzed earlier to the authors' best knowledge.  Exploiting the majorization-minimization framework \cite{hunter2004tutorial}, we have the following proposition that establishes the converging property of the fixed-point iterations in (\ref{SKEiter}).

\begin{theorem}
\label{ConvergenceMain}
 The fixed-point 
iterations in (\ref{SKEiter}) converge to the solution of (\ref{fixpoint2}) for arbitrary positive-definite initial matrices $\widehat{\mathbf{R}}_{\mathit{st}}^{(0)}$ and $\widehat{\mathbf{R}}_{p}^{(0)}$ when the conditions in \textit{Proposition} \ref{ExistenceMain} are satisfied. 
\end{theorem}

 {\textit{Proof}: See Appendix \ref{Convergence}.}

\begin{remark} The iterations in (\ref{SKEiter}) can be terminated by using a distance metric 
\begin{equation}
    \begin{split}
\mathcal{D}(\widehat{\mathbf{R}}^{(k+1)},\widehat{\mathbf{R}}^{(k)})={\left\Vert  \frac{\widehat{\mathbf{R}}^{(k+1)}}{\mathrm{Tr} (\widehat{\mathbf{R}}^{(k+1)}) }-\frac{\widehat{\mathbf{R}}^{(k)} }{\mathrm{Tr} (\widehat{\mathbf{R}}^{(k)})  } \right\Vert},
\end{split}
  \label{DistanceDefine}
\end{equation} 
where $\widehat{\mathbf{R}}^{(k)}=\widehat{\mathbf{R}}_{\mathit{st}}^{(k)} \otimes \widehat{\mathbf{R}}_p^{(k)}$ and $\Vert \cdot \Vert$ denotes the Frobenius norm. This metric measures the variation of the solution over iterations. 
Then a stopping criterion can be set to terminate the iterations when  
\begin{equation}
    \begin{split}
\mathcal{D}(\widehat{\mathbf{R}}^{(k+1)},\widehat{\mathbf{R}}^{(k)})<\delta 
\end{split}
  \label{stop}
\end{equation} 
or $k>K_{\max}$ is met, where $\delta$ denotes a preset threshold and $K_{\max}$ the maximum number of iterations allowed.

\end{remark}

\section{Choice of the Shrinkage Factors}
The performance of the RSKE depends highly on the choice of the shrinkage factors $\rho_{\mathit{st}}$ and $\rho_p$.  In practice, however, the optimal shrinkage factors are unavailable since the true CM is unknown.
In this section, we propose two different choices, based on oracle approximating shrinkage (OAS) and leave-one-out cross validation (LOOCV), respectively, to provide solutions with different performance and complexity.

\subsection{The KOAS Method}
In \cite{5743027}, an OAS strategy for choosing the shrinkage factor for unstructured CM is derived by exploiting the MMSE criterion and plug-in estimates. We can extend this strategy to the RSKE. 
The choice of the two shrinkage factors will be decoupled into separate problems to enable a low-complexity solution. Following \cite{5743027}, we begin by assuming that the true CM $\mathbf{R}_{\mathit{st}}$ and $\mathbf{R}_p$ are already ``known''.  
Then, we choose  the shrinkage factors  $(\rho_{\mathit{st}},\rho_p)$ that achieve the MMSE of the covariance matrix estimates as
\begin{equation}
    \begin{split}
\min_{\rho_{\mathit{st}}} & \quad \mathbb{E} \left\{ \left\Vert \widehat{\mathbf{R}}_{\mathit{st}}-{\mathbf{R}}_{\mathit{st}} \right\Vert^2 \right\}\\
\mathrm{s.t.} & \quad \widehat{\mathbf{R}}_{\mathit{st}}
=(1-\rho_{\mathit{st}})\mathbf{C}_{\mathit{st}}+\rho_{\mathit{st}}\mathbf{I}_{N_{\mathit{st}}},\\
    \end{split}
    \label{rhoAequation}
\end{equation} 
and
\begin{equation}
    \begin{split}
\min_{\rho_p} & \quad \mathbb{E} \left\{ \left\Vert \widehat{\mathbf{R}}_p-{\mathbf{R}}_p \right\Vert^2 \right\}\\
\mathrm{s.t.} & \quad \widehat{\mathbf{R}}_p
=(1-\rho_p)\mathbf{C}_p+\rho_p\mathbf{I}_{N_p},\\
    \end{split}
    \label{rhoBequation}
\end{equation} 
where  $\mathbb{E} \{ \cdot \}$ denotes the mathematical expectation and
\begin{equation}
    \begin{split}
\mathbf{C}_{\mathit{st}}\triangleq \frac{N}{ L N_p} \sum_{l=1}^L \frac{\mathbf{Y}_l^\mathrm{H} \mathbf{R}_p^{-1} \mathbf{Y}_l}{\mathbf{y}_l^\mathrm{H} \left( \mathbf{R}_{\mathit{st}} \otimes \mathbf{R}_p \right)^{-1}  \mathbf{y}_l}, \\
\mathbf{C}_p\triangleq \frac{N}{ L N_{\mathit{st}}} \sum_{l=1}^L \frac{\mathbf{Y}_l \mathbf{R}_{\mathit{st}}^{-1} \mathbf{Y}_l^\mathrm{H}}{\mathbf{y}_l^\mathrm{H} \left( \mathbf{R}_{\mathit{st}} \otimes \mathbf{R}_p \right)^{-1}  \mathbf{y}_l}.
    \end{split}
\end{equation} 
The following proposition extends the OAS solution of \cite{5743027}  to the Kronecker-structured CM.

\begin{theorem}
\label{ProKOAS}
The shrinkage factors that achieve the MMSE are given as (\ref{rhoAopt0}) and (\ref{rhoBopt0}) in the following page.
\begin{figure*}[!t]
\normalsize
\begin{subequations}
\label{rhoopt}
\begin{equation}
    \begin{split}
\rho_{\textit{st}}^\star=\frac{\mathrm{Tr}^2(\mathbf{R}_{\textit{st}}) -\frac{1}{N_{\textit{st}}}\mathrm{Tr}\left( {\mathbf{R}}_{\textit{st}}^2 \right)}{\left(\mathrm{Tr}^2(\mathbf{R}_{\textit{st}})+\left( 1-\frac{2\mathrm{Tr}(\mathbf{R}_{\textit{st}})}{N_{\textit{st}}} \right)  (LN+L) \right)+\left(N_p L+\frac{L-1}{N_{\textit{st}}} \right)\mathrm{Tr}\left(  {\mathbf{R}}_{\textit{st}}^2\right)},
\end{split}
\label{rhoAopt0}
\end{equation}
\begin{equation}
    \begin{split}
\rho_p^\star=\frac{\mathrm{Tr}^2(\mathbf{R}_p) -\frac{1}{N_p}\mathrm{Tr}\left( {\mathbf{R}}_p^2\right)}{\left(\mathrm{Tr}^2(\mathbf{R}_p)+\left( 1-\frac{2\mathrm{Tr}(\mathbf{R}_p)}{N_p} \right)  (LN+L) \right)+\left(N_{\textit{st}} L+\frac{L-1}{N_p} \right)\mathrm{Tr}\left(  {\mathbf{R}}_p^2 \right)}.
\end{split}
\label{rhoBopt0}
\end{equation}
\end{subequations}
\hrulefill
\vspace*{4pt}
\end{figure*}
\end{theorem}

\textit{Proof}:  See Appendix \ref{proofPro1}.

In practice, $\mathbf{R}_{\mathit{st}}$ and $\mathbf{R}_p$ in (\ref{rhoopt}) are unknown. 
Similarly to \cite{5743027}, we propose to replace them by their trace-normalized estimates $\widetilde{\mathbf{R}}_{\mathit{st}}$ and $\widetilde{\mathbf{R}}_p$, such as the KNSCM \cite{7950961} and KMLE \cite{7439863}. 
We will show the performance of the resulting shrinkage factors $(\rho_{\textit{st}, \mathrm{KOAS}}, \rho_{p, \mathrm{KOAS}})$, referred to as the Kronecker OAS (KOAS) choice, in Section IV. 
Note that, if $N_{\mathit{st}} = 1$ or $N_p = 1$, the Kronecker-structured CM reduces to the unstructured CM and (\ref{rhoopt}) agrees with (17)  in \cite{5743027}. 
If $\rho_{\textit{st}, \mathrm{KOAS}} <0$ is produced, we then truncate it to  $\rho_{\textit{st}, \mathrm{ KOAS}} =  0$. If $\rho_{\textit{st}, \mathrm{KOAS}}\ge 1$, we simply set the covariance matrix estimate to be the shrinkage target matrix. The treatments are similar for $\rho_{p, \mathrm{KOAS}} <0$ and $\rho_{p, \mathrm{KOAS}} \ge 1$ and also the LOOCV-based choices of the shrinkage factors to be introduced in the next subsection.

\subsection{The LOOCV Method} 
We next provide an alternative  for choosing the shrinkage factors based on LOOCV.
In order to achieve good performance and complexity tradeoff, the cost for LOOCV must be carefully chosen. 
In this work, we extend the quadratic cost used in \cite{TONG2018223} to obtain  a data-driven, analytical solution. 
Note that \cite{TONG2018223} considers unstructured CM for Gaussian data, whereas this paper considers Kronecker-structured CM estimation with elliptically distributed data for which iterative solvers are required. 

Let $\mathbf{\Sigma}_{\mathit{st}}$ and $\mathbf{\Sigma}_p$ be two positive-definite, Hermitian matrices. Define the following cost function
\begin{subequations}
\label{expectcostAB}
\begin{equation}
    \begin{split}
\mathcal{J}_{\mathit{st}}\left( \mathbf{\Sigma}_{\mathit{st}}\right)=\mathbb{E} \left( \left\Vert \mathbf{\Sigma}_{\mathit{st}}-  \mathbf{S}_{\mathit{st}} \right\Vert^2     \right),
    \end{split}
    \label{expectcostA}
\end{equation} 
\begin{equation}
    \begin{split}
\mathcal{J}_{p}\left( \mathbf{\Sigma}_p\right)=\mathbb{E} \left(\left\Vert \mathbf{\Sigma}_p  -  \mathbf{S}_p \right\Vert^2     \right), 
    \end{split}
    \label{expectcostB}
\end{equation} 
\end{subequations}
where the expectation is with respect to $\mathbf Y=\mathrm{unvec}_{N_p N_{\mathit{st}}}(\mathbf{y})$, 
\begin{equation}
    \begin{split}
\mathbf{S}_{\mathit{st}}\triangleq \frac{N_{\mathit{st}} \mathbf{Y}^\mathrm{H} \mathbf{R}_p^{-1} \mathbf{Y}}{\mathbf{y}^\mathrm{H} \left( \mathbf{R}_{\mathit{st}} \otimes \mathbf{R}_p \right)^{-1}  \mathbf{y}}, \mathbf{S}_p \triangleq \frac{N_p \mathbf{Y} \mathbf{R}_{\mathit{st}}^{-1} \mathbf{Y}^\mathrm{H}}{\mathbf{y}^\mathrm{H} \left( \mathbf{R}_{\mathit{st}}^{-1} \otimes \mathbf{R}_{p}^{-1} \right)  \mathbf{y}}.
    \end{split}
    \label{expectSAB}
\end{equation} 
\begin{theorem}
\label{ExpectMain}
The expectation of $\mathbf{S}_{\mathit{st}}$ and $\mathbf{S}_p$ are respectively given as $\mathbb{E}\left(\mathbf{S}_{\mathit{st}}\right)=\mathbf{R}_{\mathit{st}}$ and $\mathbb{E}\left(\mathbf{S}_p\right)=\mathbf{R}_p$,
and $\mathcal{J}_{\mathit{st}}\left( \mathbf{\Sigma}_{\mathit{st}}\right)$ and $\mathcal{J}_{p}\left( \mathbf{\Sigma}_p\right)$ are minimized by   
$\mathbf{\Sigma}_{\mathit{st}}={\mathbf{R}}_{\mathit{st}}$ and $\mathbf{\Sigma}_p={\mathbf{R}}_p$, respectively. 
\end{theorem}

\textit{Proof}: See Appendix \ref{Expect}.

Inspired by \emph{Proposition} \ref{ExpectMain}, we aim to estimate the cost function in (\ref{expectcostAB}) and then minimize it over the shrinkage factors. This may be achieved using different strategies, e.g., \cite{LEDOIT2004365}. In this paper, we apply the LOOCV strategy \cite{arlot2010survey} to estimate $\mathcal{J}_{\mathit{st}}\left( \mathbf{\Sigma}_{\mathit{st}}\right)$ and $\mathcal{J}_{p}\left( \mathbf{\Sigma}_p\right)$ and minimize them to determine the shrinkage factors. With the standard LOOCV, the samples $\mathcal{Y}$ are repeatedly split into two sets. For the $l$th split, the samples in the training set $\mathcal{Y}_l$ (with the $l$th sample $\mathbf{y}_l$ omitted from $\mathcal{Y}$) are used for producing shrinkage CM estimates $\{ \mathbf{\Sigma}_{\mathit{st}}, \mathbf{\Sigma}_p \}$ and the remaining sample $\mathbf{y}_l$ is used for constructing $\{ \mathbf{S}_{\mathit{st}}, \mathbf{S}_p \}$ to estimate $\mathcal{J}_{\mathit{st}}\left( \mathbf{\Sigma}_{\mathit{st}}\right)$ and $\mathcal{J}_{p}\left( \mathbf{\Sigma}_p\right)$.
The standard LOOCV process requires the iterative estimator to be applied for $L$ times for each pair of candidate shrinkage factors $(\rho_{\mathit{st}}, \rho_p)$, which can lead to significant complexity, especially when grid search of $(\rho_{\mathit{st}}, \rho_p)$ is conducted. In order to address this complexity challenge, we propose an alternative solution  by using proxy estimators so that closed-form expressions can be found for the optimized shrinkage factors. 

Similarly to KOAS, we first assume that the covariance matrices are ``known'' and consider  estimates of the covariance matrices from the samples $\mathcal{Y}_l=\{\mathbf Y_j, j\ne l\}$ as    
\begin{subequations}
\begin{equation}
    \begin{split}
\widehat{\mathbf{R}}_{\mathit{st}}^{(l)} (\rho_{\mathit{st}})&=(1-\rho_{\mathit{st}})\widehat{\mathbf{C}}_{\mathit{st}}^{(l)}+\rho_{\mathit{st}}\mathbf{I}_{N_{\mathit{st}}},\\
\end{split}
\label{CVestA}
\end{equation} 
\begin{equation}
    \begin{split}
\widehat{\mathbf{R}}_p^{(l)} (\rho_p)&=(1-\rho_p)\widehat{\mathbf{C}}_p^{(l)}+\rho_p\mathbf{I}_{N_p},\\
\end{split}
\label{CVestB}
\end{equation} 
\end{subequations}
where
\begin{subequations}
\begin{equation}
    \begin{split}
\widehat{\mathbf{C}}_{\mathit{st}}^{(l)}&=\frac{N_{\mathit{st}}}{ L-1} \sum_{j\neq l} \frac{\mathbf{Y}_j^\mathrm{H} \mathbf{R}_p^{-1} \mathbf{Y}_j}{\mathbf{y}_j^\mathrm{H} \left( {\mathbf{R}}_{\mathit{st}} \otimes {\mathbf{R}}_p \right)^{-1}  \mathbf{y}_j},\\
\end{split}
\label{CAk1l}
\end{equation} 
\begin{equation}
    \begin{split}
\widehat{\mathbf{C}}_p^{(l)}=\frac{N_p}{L-1} \sum_{j\neq l} 
\frac{\mathbf{Y}_j \mathbf{R}_{\mathit{st}}^{-1} \mathbf{Y}_j^\mathrm{H}}{\mathbf{y}_j^\mathrm{H} \left( {\mathbf{R}}_{\mathit{st}}\otimes {\mathbf{R}}_p \right)^{-1}  \mathbf{y}_j}.
\end{split}
\label{CBk1l}
\end{equation} 
\end{subequations}
Following \cite{TONG2018223}, we adopt the quadratic cost functions below:
\begin{subequations}
\begin{equation}
    \begin{split}
\mathcal{J}_{\textit{st},\mathrm{CV}}\left( \widehat{\mathbf{R}}_{\mathit{st}}\right)=\frac{1}{L}\sum_{l=1}^L \left\Vert \widehat{\mathbf{R}}_{\mathit{st}}^{(l)} (\rho_{\mathit{st}}) - \widehat{\mathbf{S}}_{\mathit{st}}^{(l)} \right\Vert^2,
    \end{split}
    \label{LOOCVcostA}
\end{equation} 
\begin{equation}
    \begin{split}
\mathcal{J}_{p,\mathrm{CV}}\left( \widehat{\mathbf{R}}_p\right)=\frac{1}{L}\sum_{l=1}^L \left\Vert \widehat{\mathbf{R}}_p^{(l)} (\rho_p) - \widehat{\mathbf{S}}_p^{(l)} \right\Vert^2,
    \end{split}
    \label{LOOCVcostB}
\end{equation}
\end{subequations} 
where
\begin{subequations}
    \label{LOOCVS}
\begin{equation}
    \begin{split}
\widehat{\mathbf{S}}_{\mathit{st}}^{(l)}=\frac{N_{\mathit{st}} \mathbf{Y}_l^\mathrm{H} \mathbf{R}_p^{-1} \mathbf{Y}_l}{\mathbf{y}_l^\mathrm{H} \left( {\mathbf{R}}_{\mathit{st}} \otimes {\mathbf{R}}_p \right)^{-1}  \mathbf{y}_l},
    \end{split}
    \label{LOOCVSA}
\end{equation} 
\begin{equation}
    \begin{split}
\widehat{\mathbf{S}}_p^{(l)}= \frac{N_p \mathbf{Y}_l \mathbf{R}_{\mathit{st}}^{-1} \mathbf{Y}_l^\mathrm{H}}{\mathbf{y}_l^\mathrm{H} \left( {\mathbf{R}}_{\mathit{st}}\otimes {\mathbf{R}}_p \right)^{-1}  \mathbf{y}_l}.
    \end{split}
    \label{LOOCVSB}
\end{equation} 
\end{subequations}
Substituting (\ref{CVestA}) into (\ref{LOOCVcostA}), the cost function can be rewritten as
\begin{equation}
    \begin{split}
\mathcal{J}_{\textit{st},\mathrm{CV}}\left( \rho_{\mathit{st}}\right)&=\frac{1}{L}\sum_{l=1}^L \left\Vert (1-\rho_{\mathit{st}}) \widehat{\mathbf{C}}_{\mathit{st}}^{(l)} +\rho_{\mathit{st}} \mathbf{I}_{N_{\mathit{st}}}-  \widehat{\mathbf{S}}_{\mathit{st}}^{(l)} \right\Vert^2.\\
    \end{split}
    \label{LOOCVcost2A}
\end{equation} 
We treat $\mathcal{J}_{\textit{st},\mathrm{CV}}\left( \rho_{\mathit{st}}\right)$ as a proxy of 
$\mathcal{J}_{\mathit{st}}\left( \mathbf{\Sigma}_{\mathit{st}}\right)$  and choose the shrinkage factor $\rho_{\mathit{st}}$ as the minimizer of (\ref{LOOCVcost2A}) as:
\begin{equation}
    \begin{split}
&\rho_{\textit{st},\mathrm{CV}} =\frac{\mathrm{Re} \left( \sum\limits_{l=1}^L\mathrm{Tr}  \left[  \left(\mathbf{I}_{N_{\mathit{st}}}- \widehat{\mathbf{C}}_{\mathit{st}}^{(l)} \right) \left(\widehat{\mathbf{S}}_{\mathit{st}}^{(l)}- \widehat{\mathbf{C}}_{\mathit{st}}^{(l)} \right)  \right] \right)}{ \sum\limits_{l=1}^L \mathrm{Tr}  \left[  \left(\mathbf{I}_{N_{\mathit{st}}}- \widehat{\mathbf{C}}_{\mathit{st}}^{(l)} \right)^2
 \right]}.\\
    \end{split}
    \label{rhoAcv}
\end{equation} 
Similarly, we choose $\rho_p$ as
\begin{equation}
    \begin{split}
&\rho_{p,\mathrm{CV}} =\frac{\mathrm{Re} \left( \sum\limits_{l=1}^L\mathrm{Tr}  \left[  \left(\mathbf{I}_{N_p}- \widehat{\mathbf{C}}_p^{(l)} \right) \left(\widehat{\mathbf{S}}_p^{(l)}- \widehat{\mathbf{C}}_p^{(l)} \right)    \right] \right)}{ \sum\limits_{l=1}^L \mathrm{Tr}  \left[   \left(\mathbf{I}_{N_p}- \widehat{\mathbf{C}}_p^{(l)} \right)^2
\right]}.
    \end{split}
    \label{rhoBcv}
\end{equation}

Alternative expressions can be derived for (\ref{rhoAcv}) and (\ref{rhoBcv}) to reduce the computational costs. 
Let
\begin{equation}
    \begin{split}
\widehat{\mathbf{C}}_{\mathit{st}}&=\frac{N_{\mathit{st}}}{ L} \sum_{l=1}^L \frac{\mathbf{Y}_l^\mathrm{H} \mathbf{R}_p^{-1} \mathbf{Y}_l}{\mathbf{y}_l^\mathrm{H} \left( {\mathbf{R}}_{\mathit{st}} \otimes {\mathbf{R}}_p \right)^{-1}  \mathbf{y}_l}.\\
\end{split}
\label{CA1l}
\end{equation} 
Recalling (\ref{CAk1l}) and (\ref{LOOCVSA}), we have 
\begin{equation}
    \begin{split}
&\widehat{\mathbf{C}}_{\mathit{st}}^{(l)}=\frac{L}{ L-1} \widehat{\mathbf{C}}_{\mathit{st}}  -\frac{1}{L-1}   \widehat{\mathbf{S}}_{\mathit{st}}^{(l)},L \widehat{\mathbf{C}}_{\mathit{st}}= \sum_{l=1}^L \widehat{\mathbf{C}}_{\mathit{st}}^{(l)}= \sum_{l=1}^L \widehat{\mathbf{S}}_{\mathit{st}}^{(l)}.
    \end{split}
    \label{CASArelation}
\end{equation} 
Note that $\widehat{\mathbf{C}}_{\mathit{st}}$, $\widehat{\mathbf{C}}_{\mathit{st}}^{(l)}$, $\widehat{\mathbf{S}}_{\mathit{st}}^{(l)}$ and $\mathbf{I}_{N_{\mathit{st}}}$ are all Hermitian matrices.
By using (\ref{CASArelation}), we have 
\begin{subequations}
\label{subexpressions}
\begin{equation}
    \begin{split}
\sum_{l=1}^L \mathrm{Tr}  \left(\widehat{\mathbf{C}}_{\mathit{st}}^{(l)}   \widehat{\mathbf{S}}_{\mathit{st}}^{(l)} \right) 
&=\frac{L^2 }{L-1} \mathrm{Tr} \left(   \widehat{\mathbf{C}}_{\mathit{st}}^2 \right) -\frac{\sum\limits_{l=1}^L \mathrm{Tr} \left( (   \widehat{\mathbf{S}}_{\mathit{st}}^{(l)}    )^2 \right)}{L-1} ,
    \end{split}
\end{equation} 
\begin{equation}
    \begin{split}
\sum_{l=1}^L \mathrm{Tr}  \left(  
(\widehat{\mathbf{C}}_{\mathit{st}}^{(l)}   )^2 \right)
&=\frac{L^2(L-2) }{(L-1)^2} \mathrm{Tr} \left(   \widehat{\mathbf{C}}_{\mathit{st}}^2 \right) +\frac{\sum\limits_{l=1}^L \mathrm{Tr} \left( (   \widehat{\mathbf{S}}_{\mathit{st}}^{(l)}   )^2 \right)}{(L-1)^2}.
    \end{split}
\end{equation} 
\end{subequations}
Substituting (\ref{subexpressions}) into   (\ref{rhoAcv}), we obtain (\ref{rhoAcv2}) on the next page to quickly evaluate the shrinkage factors $\rho_{st,\mathrm{CV}}$. Similarly, we can obtain (\ref{rhoBcv2}) there for $\rho_{p,\mathrm{CV}}$, 
where 
\begin{equation}
    \begin{split}
\widehat{\mathbf{C}}_p =\frac{N_p}{L} \sum_{l=1}^L  
\frac{\mathbf{Y}_j \mathbf{R}_{\mathit{st}}^{-1} \mathbf{Y}_j^\mathrm{H}}{\mathbf{y}_j^\mathrm{H} \left( {\mathbf{R}}_{\mathit{st}}\otimes {\mathbf{R}}_p \right)^{-1}  \mathbf{y}_j}.
\end{split}
\label{CB}
\end{equation} 
\begin{figure*}[!t]
\normalsize
\begin{subequations}
\label{rhocvone}
\begin{equation}
    \begin{split}
\rho_{\textit{st}, \mathrm{CV}}&=\frac{-\frac{L}{(L-1)^2} \mathrm{Tr} \left(   \widehat{\mathbf{C}}_{\mathit{st}}^2 \right)
+\frac{1}{(L-1)^2} \sum\limits_{l=1}^L \mathrm{Tr} \left(   ( \widehat{\mathbf{S}}_{\mathit{st}}^{(l)}  )^2 \right)
}{   N_{\mathit{st}}   -2\mathrm{Tr}  \left(       \widehat{\mathbf{C}}_{\mathit{st}} \right) +\frac{L(L-2) }{(L-1)^2} \mathrm{Tr} \left(   \widehat{\mathbf{C}}_{\mathit{st}}^2 \right)+\frac{1}{L(L-1)^2}\sum\limits_{l=1}^L \mathrm{Tr} \left( (   \widehat{\mathbf{S}}_{\mathit{st}}^{(l)}    )^2 \right).     }
    \end{split}
    \label{rhoAcv2}
\end{equation} 
\begin{equation}
    \begin{split}
\rho_{p, \mathrm{CV}} &=\frac{-\frac{L}{(L-1)^2} \mathrm{Tr} \left( \widehat{\mathbf{C}}_p^2 \right)
+\frac{1}{(L-1)^2} \sum\limits_{l=1}^L \mathrm{Tr} \left(   ( \widehat{\mathbf{S}}_p^{(l)}  )^2 \right)
}{  N_p   -2\mathrm{Tr}  \left(         \widehat{\mathbf{C}}_p \right) +\frac{L(L-2) }{(L-1)^2} \mathrm{Tr} \left(   \widehat{\mathbf{C}}_p^2 \right) +\frac{1}{L(L-1)^2}\sum\limits_{l=1}^L \mathrm{Tr} \left( (   \widehat{\mathbf{S}}_p^{(l)}    )^2 \right).     }
    \end{split}
    \label{rhoBcv2}
\end{equation} 
\end{subequations}
\hrulefill
\vspace*{4pt}
\end{figure*}

The shrinkage factors determined by (\ref{rhocvone}) 
still require the true CM $\mathbf{R}_{\mathit{st}}$ and $\mathbf{R}_p$ to be known to compute (\ref{CA1l}), (\ref{CB}), and (\ref{LOOCVS}). Similarly to KOAS, we propose to substitute them by their trace-normalized estimates  
$\widetilde{\mathbf{R}}_{\mathit{st}}$ and $\widetilde{\mathbf{R}}_p$. 
We refer to the resultant solutions as the $\mathrm{CV}$ choice.

 \begin{remark}
The proposed methods exhibit different complexities.  
If the shrinkage factors are given, the computational complexity of the iterative process in (\ref{SKEiter}) is about $\mathcal{O}(N_{it} (N_{\mathit{st}}^3 + N_p^3 + L(N_{\mathit{st}} N_p^2+N_{\mathit{st}}^2 N_p))$, where $N_{it}$ denotes the number of iterations, and we have used the identities $(\mathbf A \otimes \mathbf B)^{-1}=\mathbf A^{-1} \otimes \mathbf B^{-1}$ and $(\mathbf B^T \otimes \mathbf A) \mathrm{vec}(\mathbf X) = \mathrm{vec}(\mathbf A \mathbf X \mathbf B)$.
All the shrinkage factors proposed 
are given in closed forms without the need of grid search. Their complexities are summarized below, where only the highest order of the complexity is counted. 
\end{remark}
\begin{itemize}
\item  $\mathrm{KOAS}$: The computational complexity of (\ref{rhoopt}) mainly arises from the computation of $\mathrm{Tr}(\widetilde{\mathbf{R}}_{\mathit{st}}^2)$ and $\mathrm{Tr}(\widetilde{\mathbf{R}}_p^2)$, which is $\mathcal{O}(N_{\mathit{st}}^2 + N_p^2)$   
when the plug-in CMs  $\widetilde{\mathbf{R}}_{\mathit{st}}$ and $\widetilde{\mathbf{R}}_p$ are known. 
\item  $\mathrm{CV}$: Given $\widetilde{\mathbf{R}}_{\mathit{st}}$ and $\widetilde{\mathbf{R}}_p$, (\ref{rhocvone})  can be evaluated at a complexity of $\mathcal{O}(N_{\mathit{st}}^3 + N_p^3 + L(N_{\mathit{st}}^2 N_p + N_{\mathit{st}} N_p^2) )$.   
\end{itemize}
It can be seen that, ignoring the cost for finding the plug-in CMs, the complexity of finding the shrinkage factors is dominated by that of iteratively updating the CMs in (\ref{SKEiter}).

\section{Simulation Results}

In this section, we show the performance of the proposed RSKE estimators. 
We compare the proposed estimators with the following CM estimators: KMLE \cite{7439863,7950961}, and KNSCM \cite{7950961}. 
We will then demonstrate the superiority of our proposed methods over these existing methods with the true data and generated simulation data.

\subsection{Target detection}

\begin{figure}[!t]
\centering 
    \includegraphics[width=2.5in]{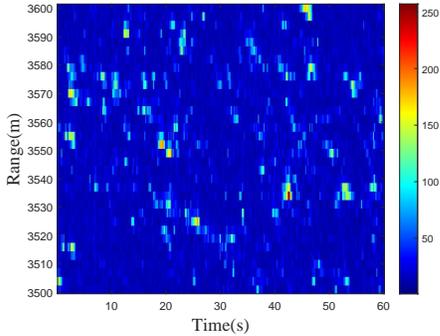}\label{fig_overview2}
\caption{The overview  of IPIX 1998 data set. }
\label{fig_overview}
\end{figure}

\begin{figure*}[!t]
\centering 
    \subfloat[]{
	\includegraphics[width=2.2in]{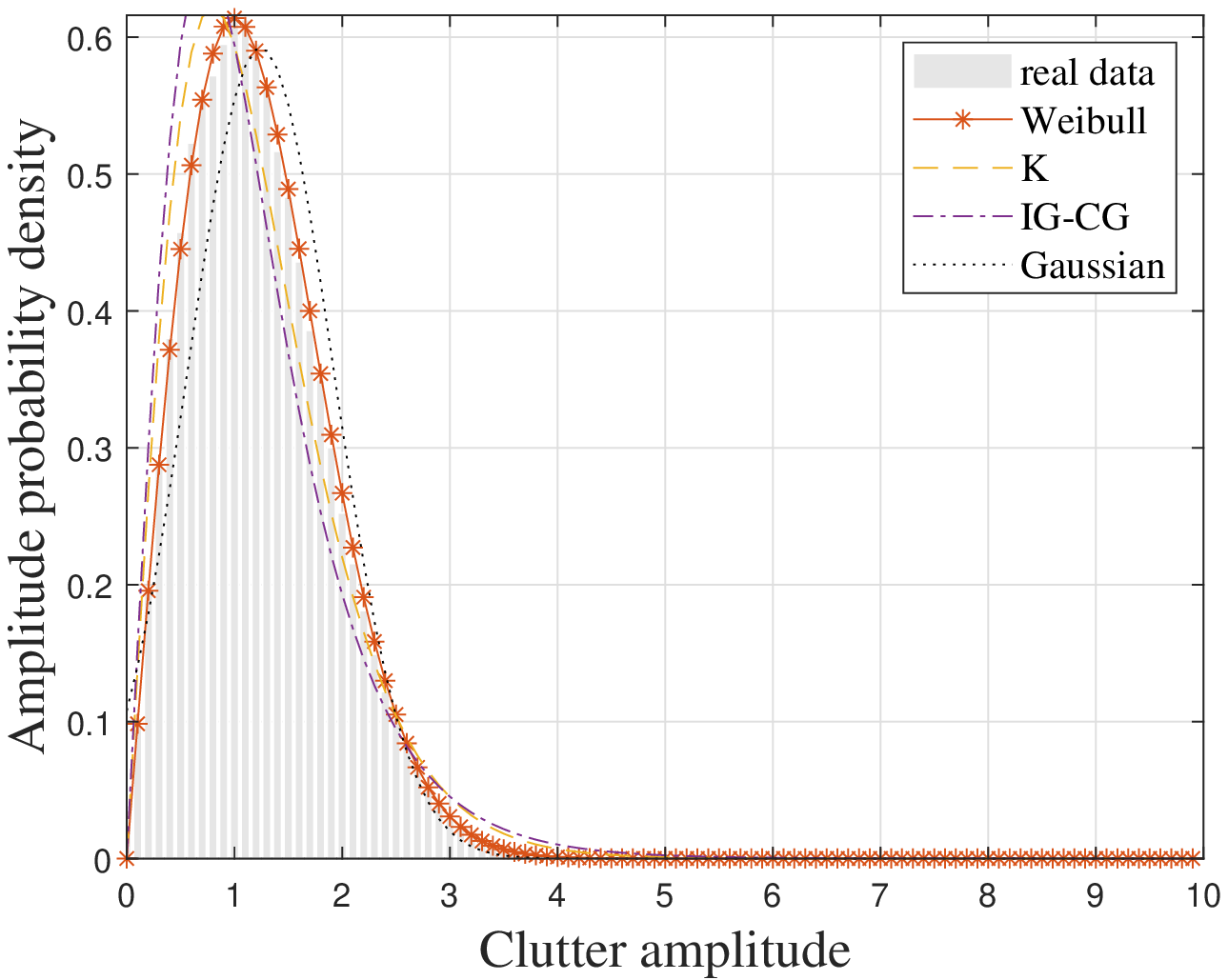}\label{fig_fitness4}}\
\subfloat[]{
	\includegraphics[width=2.2in]{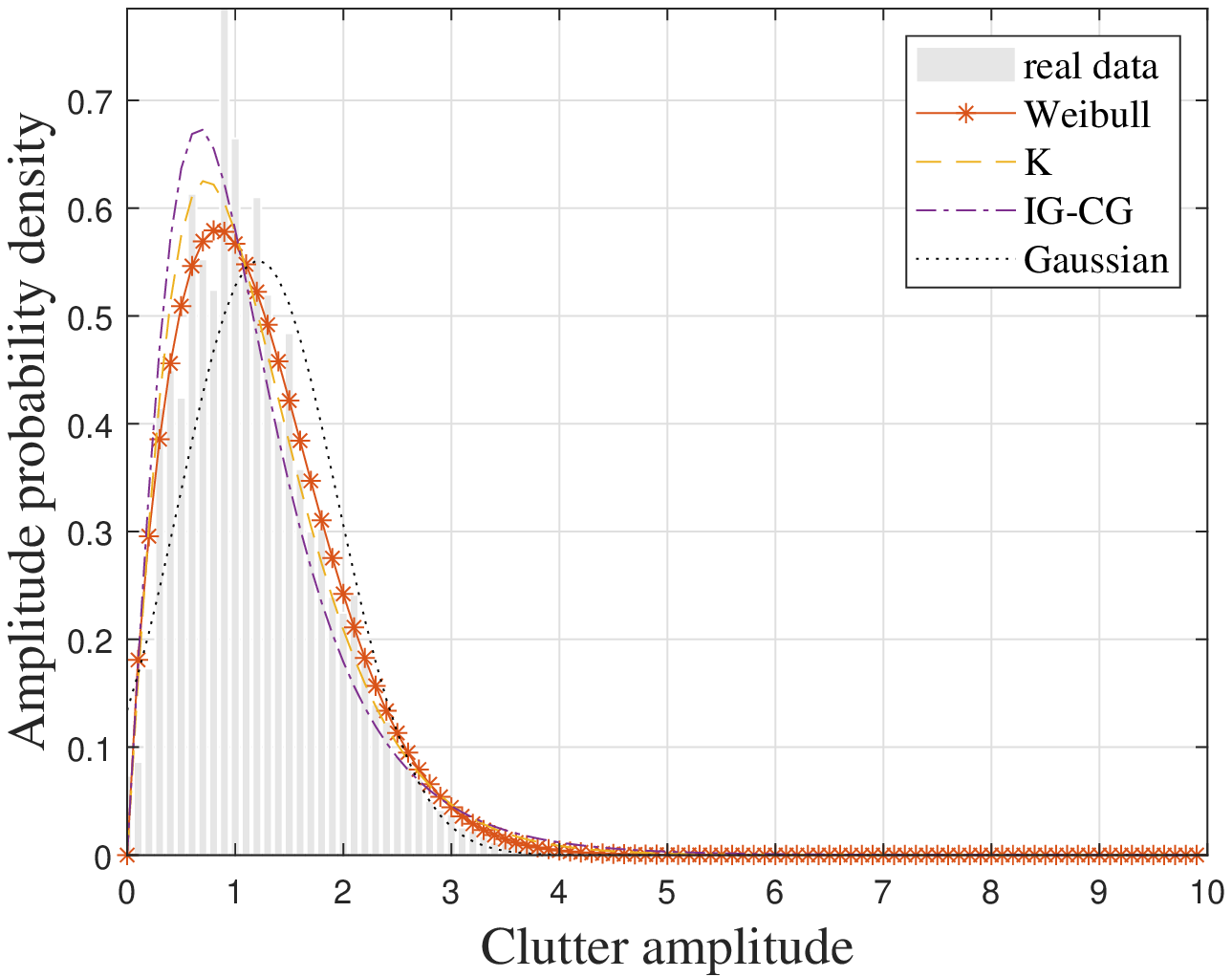}\label{fig_fitness5}}\
\subfloat[]{
	\includegraphics[width=2.2in]{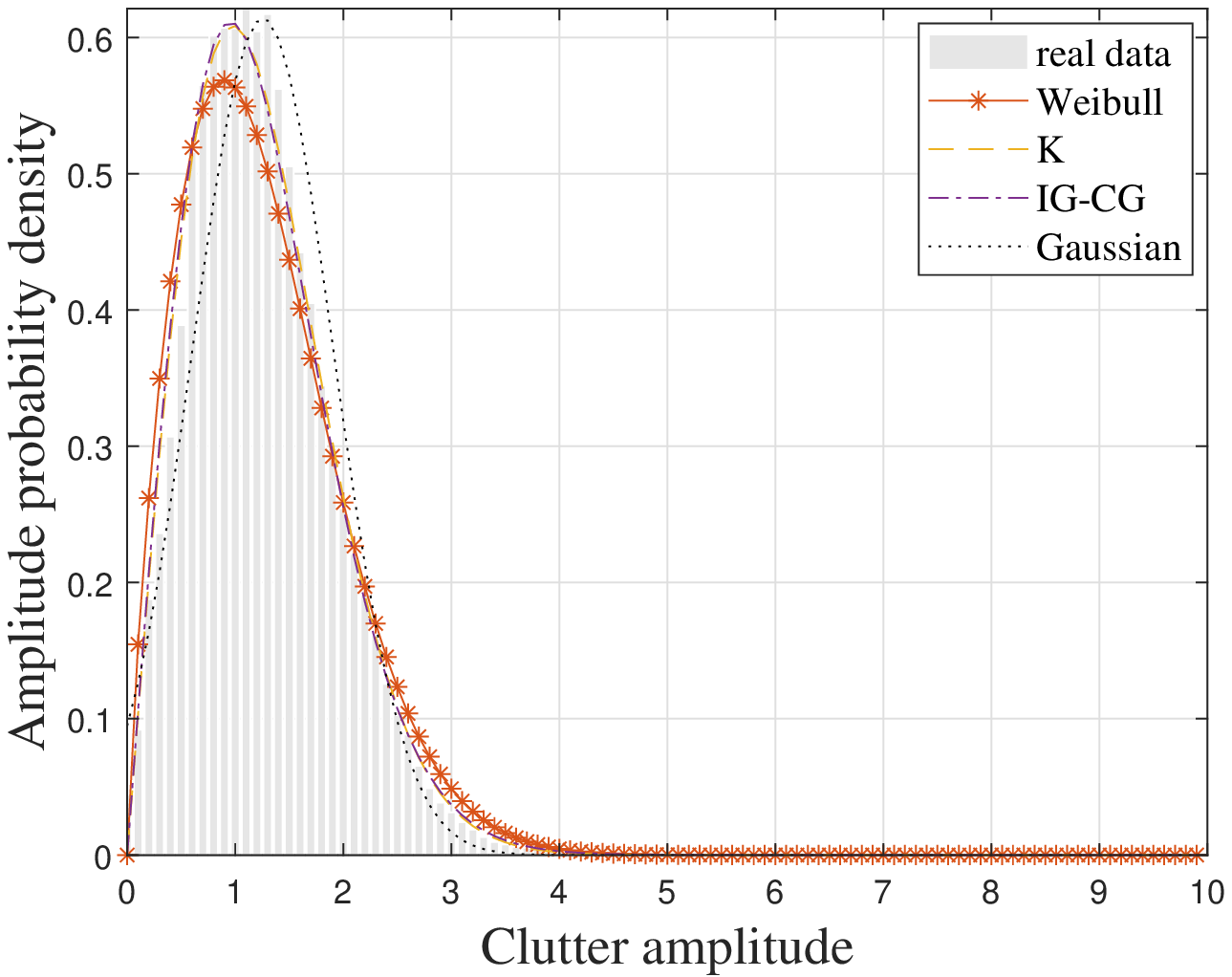}\label{fig_fitness6}}\\
    \subfloat[]{
    \includegraphics[width=2.2in]{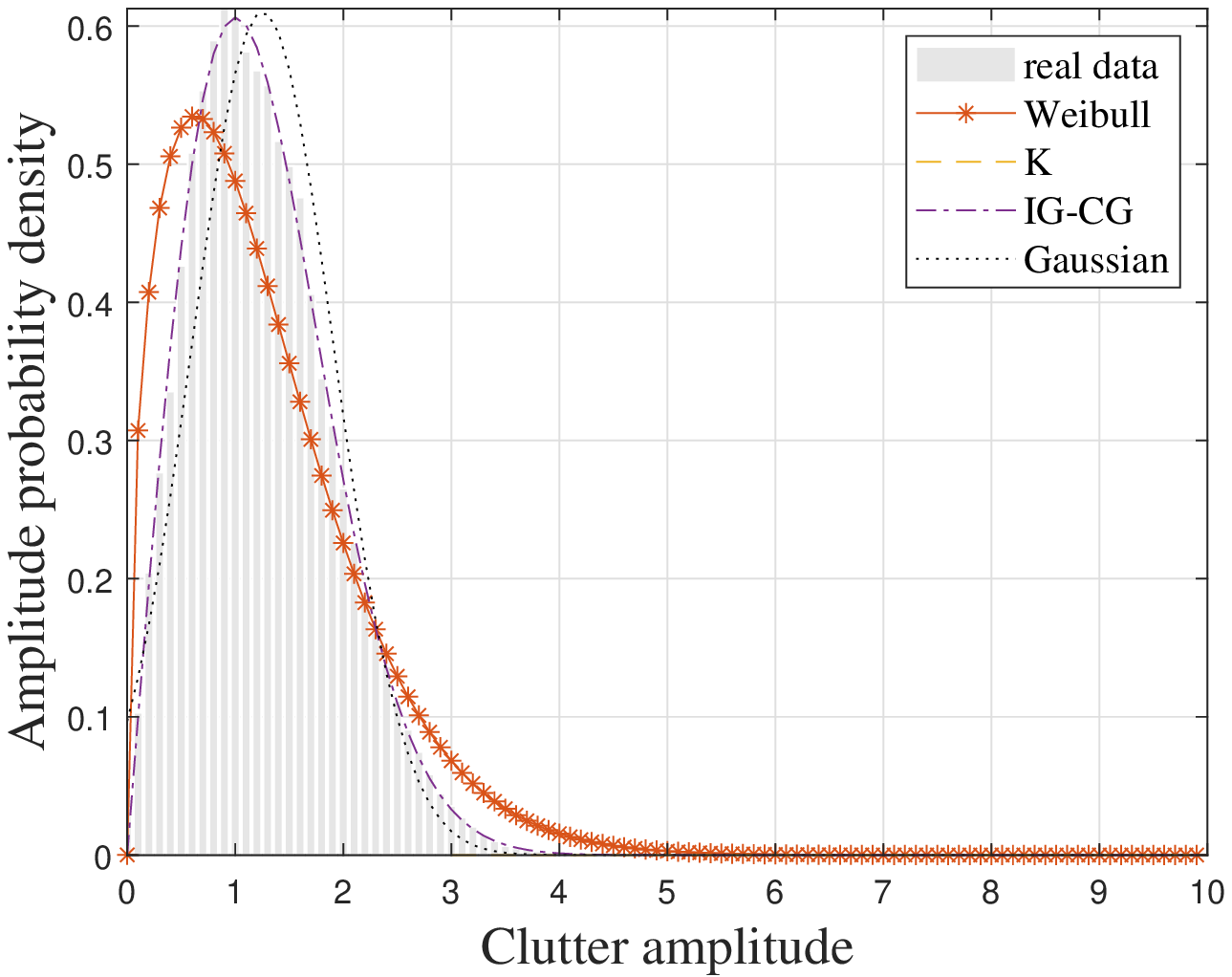}\label{fig_fitness1}}\
    \subfloat[]{
    \includegraphics[width=2.2in]{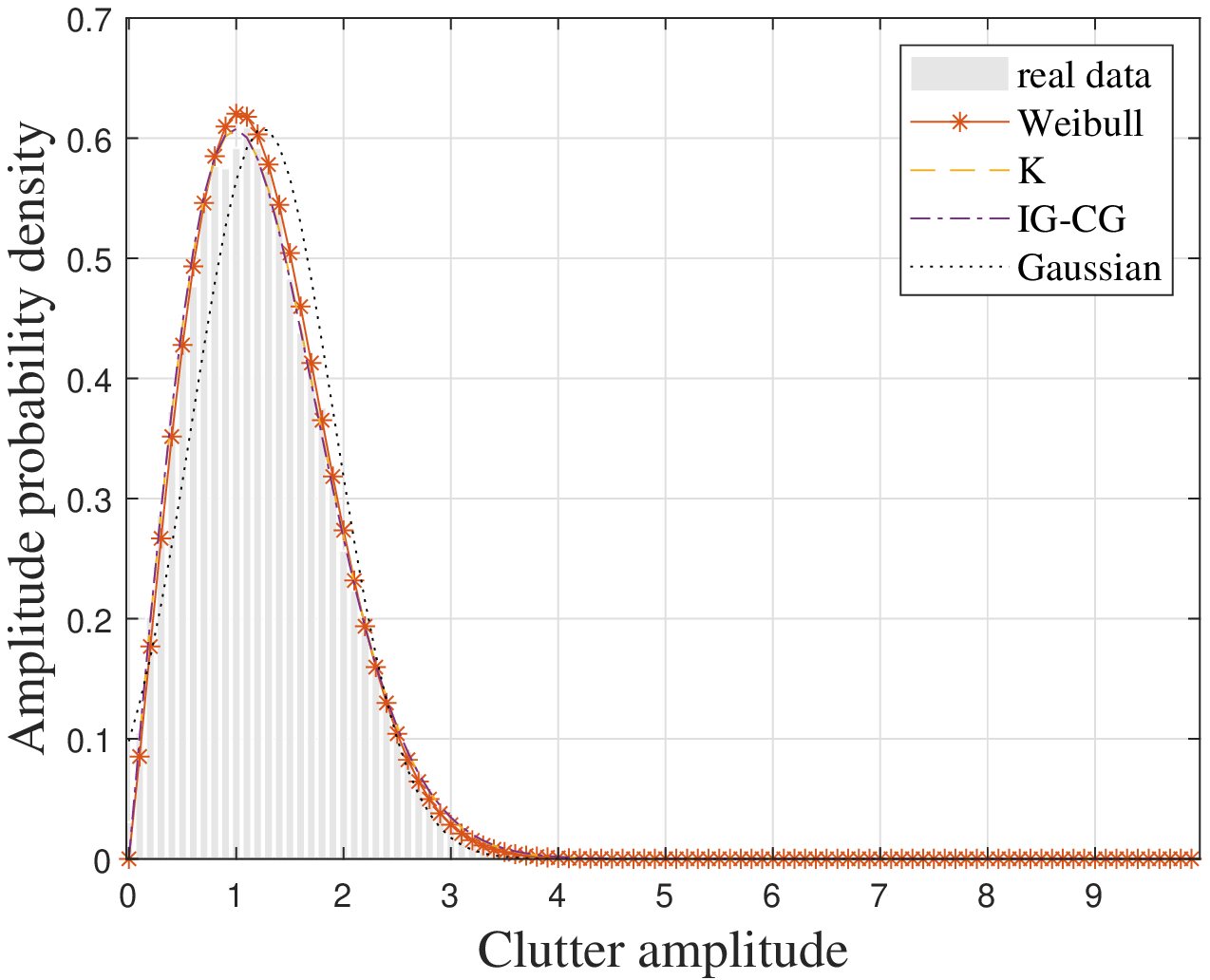}\label{fig_fitness2}}\
    \subfloat[]{
    \includegraphics[width=2.2in]{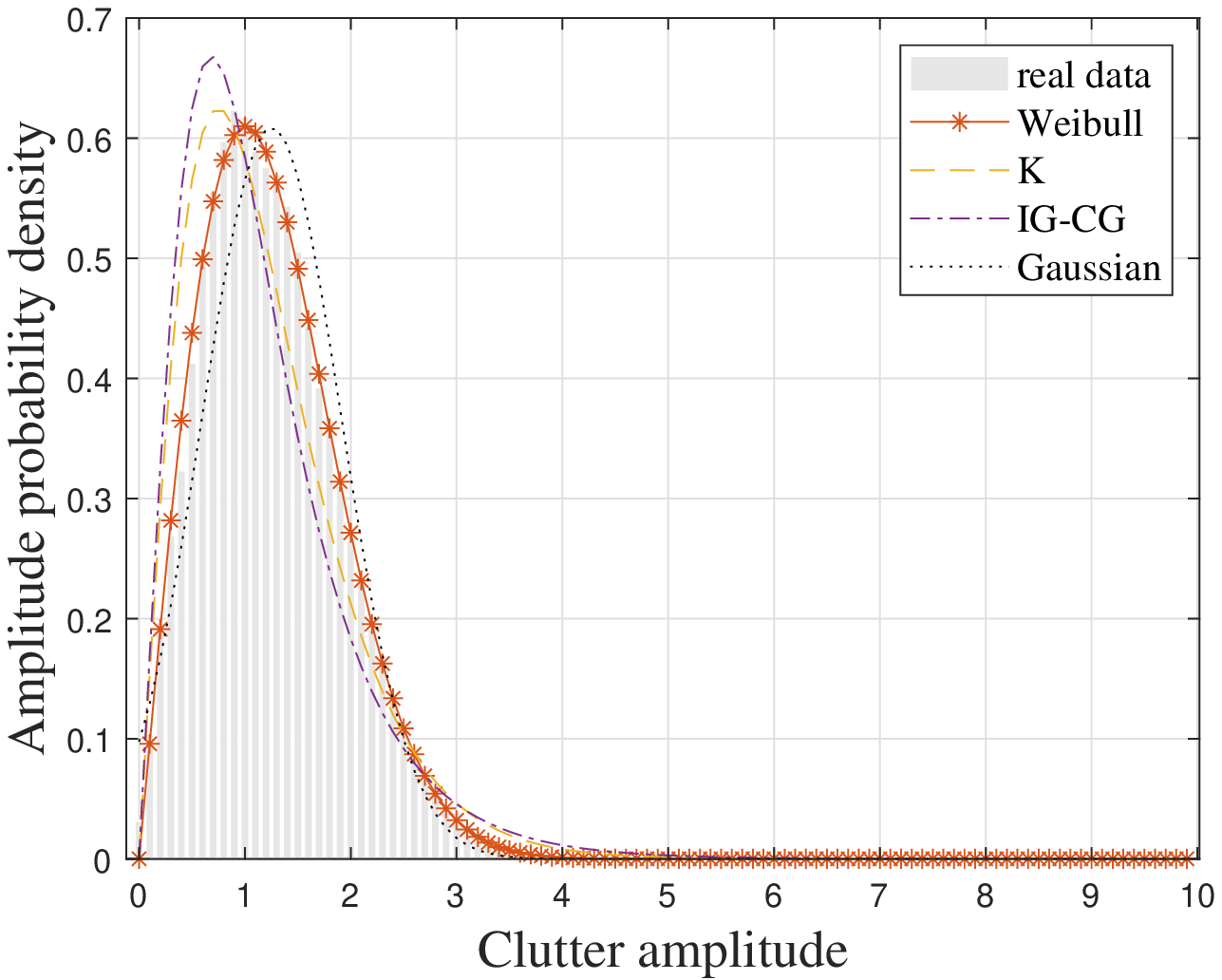}\label{fig_fitness3}}\\
\caption{Fitting the real sea clutter by Weibull, IG-CG and K distributions with different polarization. (a) HH  in 19980223$\_$171533; (b) VH  in 19980223$\_$171533;
(c) VV  in 19980223$\_$171533; (d) HH  in 19980226$\_$215015; (e) VH  in 19980226$\_$215015;
(f) VV in 19980226$\_$215015.}
\label{fig_fitness}
\end{figure*}

\begin{figure*}[!t]
\centering 
    \subfloat[]{
    \includegraphics[width=2.5in]{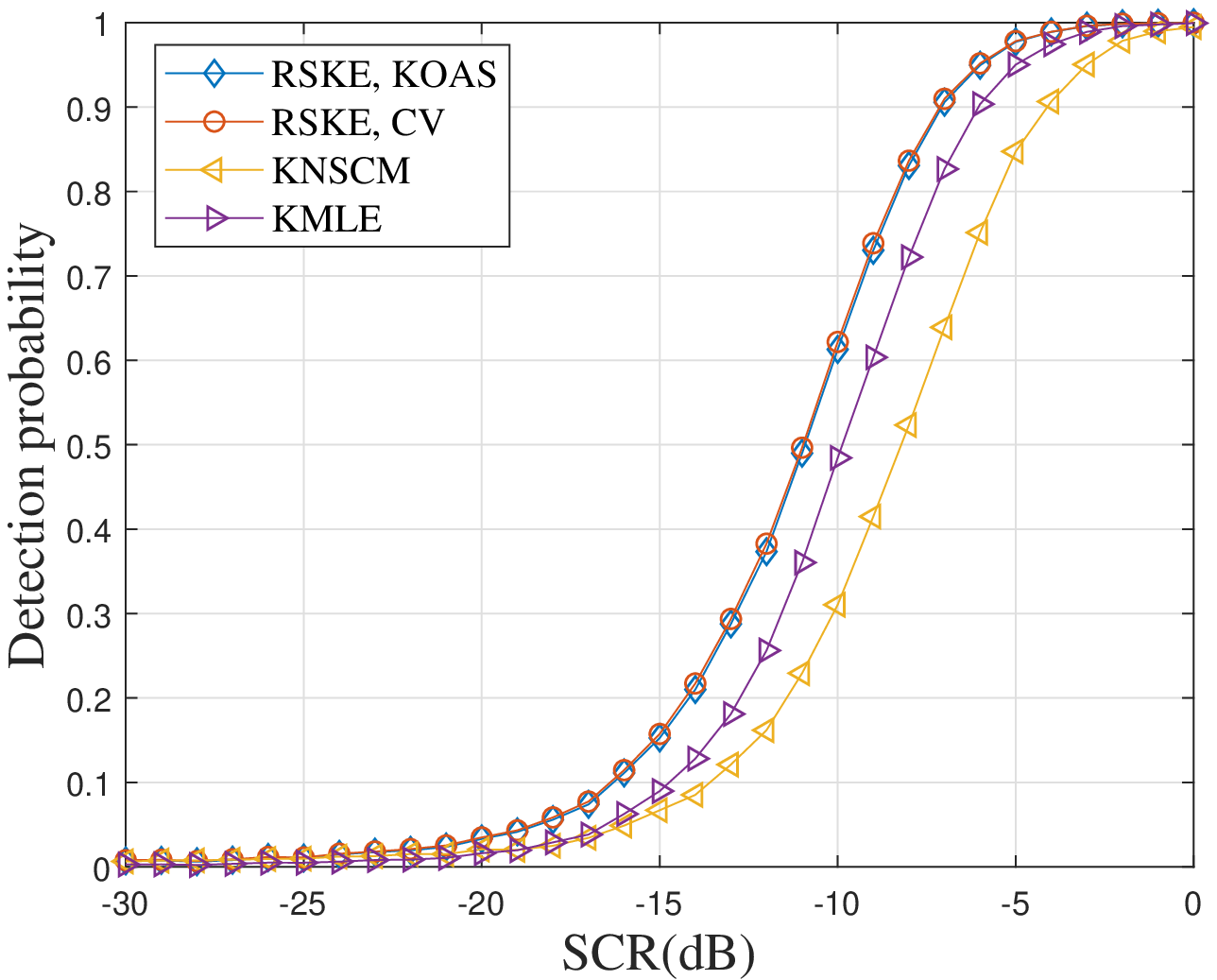}\label{fig_PDvsSCNR1}}\
    \subfloat[]{
    \includegraphics[width=2.5in]{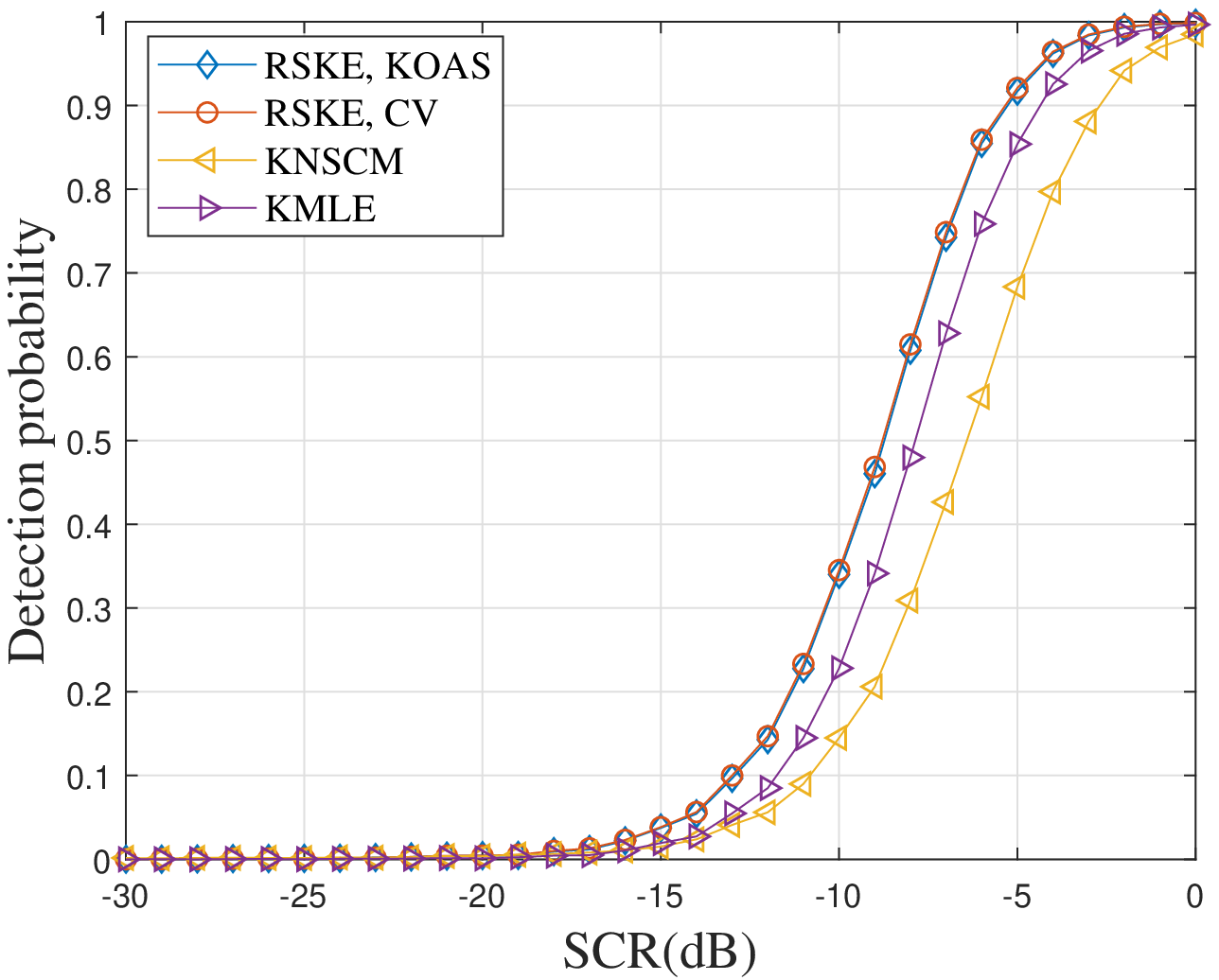}\label{fig_PDvsSCNR2}}\\
\caption{Detection performance versus the SCR. (a) $P_{fa}=10^{-2}$;
(b)  $P_{fa}=10^{-4}$.}
\label{fig_PDvsSCNR}
\end{figure*}

In this subsection, we show simulation results to demonstrate the performance of the RSKE for the polarization target detection in the context of real heterogeneous sea clutter data.
Ice Multiparameter Imaging X-Band (IPIX) 1998 is collected using the McMaster IPIX radar with one single antenna from Grimsby, Canada \cite{IPIX}.
One data set that we use is IPIX 1998 file ``19980223$\_$171533''. 
In Fig. 1, we show the normalized logarithmic amplitude of the clutter in this file.
Key parameters of the data set include the carrier frequency $9.39$GHz, PRF $1000$Hz, pulse length $20$ns and range resolution 3m. We refer the reader to the official website \cite{IPIX} for more details.
From Fig. 1, we can see that there are many strong scattering points whose echo amplitude is significantly large. This indicates that the data fit the compound Gaussian distribution better due to its heavy tail in contrast to the Gaussian one.

\begin{table}[!t]
	\renewcommand{\arraystretch}{1.3}
	\caption{Fitting error}
	\label{tableerr}
	\centering
	\begin{tabular}{|c||c|}
		\hline
		Distribution & Error ($\times 10^{-4}$)\\
		\hline
		Gaussian & $12.1012$\\
		Weibull & $9.5881$\\
		IG-CG & $3.0916$\\
		K &    $2.4125$\\
		\hline
	\end{tabular}
\end{table}

In order to illustrate this, we use the compound Gaussian distribution to fit the  probability density function of the amplitude of the sea clutter in file ``19980223$\_$171533'' and ``19980226$\_$215015'' under different polarization. Note that the data correspond to different temperatures, wind directions, wind speeds, wave heights, wave periods, precipitation, etc.
The curves for fitting the amplitude using three types of CG distribution (including the Weibull, inverse Gamma-compound Gaussian (IG-CG) and K distributions) are plotted in Fig. \ref{fig_fitness}. 
From Fig. \ref{fig_fitness}, we can see that the real sea clutter data have a heavier tail than the Gaussian model. 
The fitting errors\footnote{The fitting error is defined as the mean square error (MSE) between the empirical p.d.f. of the real data and fitting distributions.} for the VV data of ``19980223$\_$171533'' are given in Table \ref{tableerr}, which demonstrates that the fitting error of the Gaussian distribution is larger than that of the CG distributions.
This shows the suitability of the CG model for fitting the real sea clutter. Note that under different sea states, the different types of CG distribution may provide different accuracies for fitting the clutter data. However, the CG model always fits the data better than the Gaussian one. Meanwhile, the proposed RSKE is effective for various CG data, regardless of the specific type.

To  assess the detection performance, we consider the well known normalized matched filter (NMF) detector\cite{6913007}, i.e.,  
 \begin{equation}
    \begin{split}
	\Lambda= \frac{\left| \mathbf s ^\mathrm{H} \widehat{\mathbf{R}}^{-1}  \mathbf{y}   \right|^2 }{\left(\mathbf s ^\mathrm{H} \widehat{\mathbf{R}}^{-1}  \mathbf s  \right)
\left(\mathbf y ^\mathrm{H} \widehat{\mathbf{R}}^{-1} \mathbf y   \right)}  \mathop{\gtrless}\limits_{H_0}^{H_1}  \delta.
    \end{split}
\end{equation}
Recall that $\mathbf s $ denotes the steering vector of desired signal, $\widehat{\mathbf{R}}$ denotes the estimated CM, $\mathbf y$ denotes the received echo, and  $\delta$ denotes the detection threshold.

In order to obtain $\delta$, we first implement $100/P_{fa}$ Monte-Carlo trials to ensure a preassigned
value of the probability of false alarm $P_{fa}$. In this section, we set $N_s=1$, $N_t=8$, $L=8$. The normalized Doppler frequency of the  target is 0.25 and its azimuth and elevation angles are $0^\circ$ and $3.6^\circ$, respectively. 
We use three different polarization channels, i.e., HH, HV and VV.
Note that the SCR is computed as  $\mathrm{SCR} = 
\frac{   \sigma_s^2}{\sigma_c^2}$,
where $\sigma_s^2$ and $\sigma_c^2$ are the power of the target and clutter, respectively.

 Fig. \ref{fig_PDvsSCNR} shows the detection performance for the NMF versus the input SCR.
For each abscissa, 10000 Monte-Carlo experiments are performed. 
It is seen that the proposed methods can achieve the best detection performance among several estimators under different $P_{fa}$.
For example, when the SCR is $-10$ dB, the detection probability with the proposed estimators is about 62$\%$ while that with KMLE and KNSCM are 49$\%$ and 31$\%$, respectively. 
This shows that the RSKE is effective for the target detection application with a similar computational complexity as that of the KMLE.

\begin{figure*}[!t]
\centering 
    \subfloat[]{
    \includegraphics[width=2.5in]{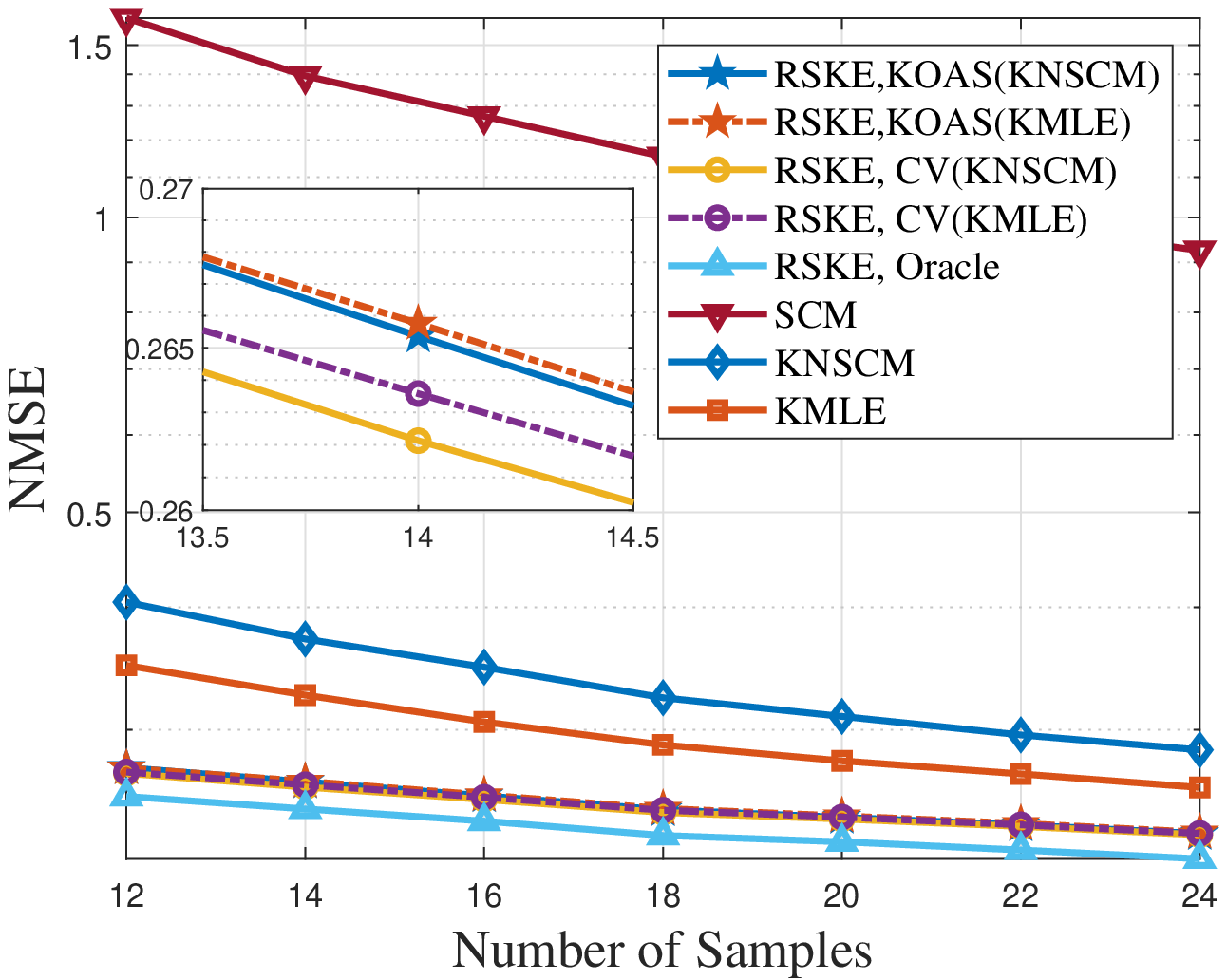}\label{fig_mse1}}\
    \subfloat[]{
    \includegraphics[width=2.5in]{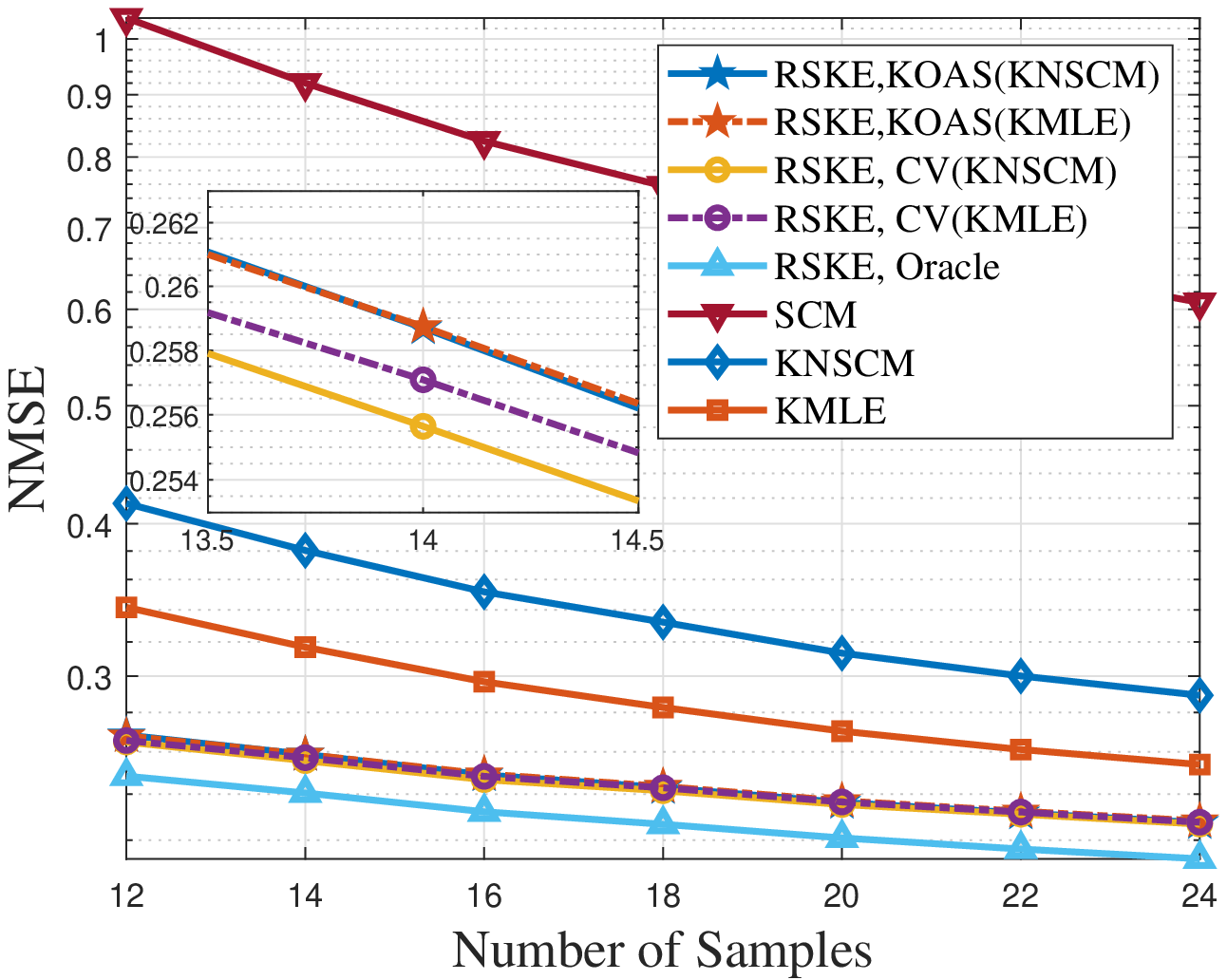}\label{fig_mse3}}\\
\caption{NMSE versus the number of samples $L$. 
(a) $\nu=1$;
(b) $\nu=10$; }
\label{fig_mse}
\end{figure*}

\subsection{CM Estimation Accuracy}
In order to evaluate the CM estimation accuracy, we use the following normalized mean-square error (NMSE) as the performance metric \cite{7539668}: 
 \begin{equation}
    \begin{split}
	\mathrm{NMSE} \triangleq  \frac{\mathbb{E}  \left\{ \left\Vert \widehat{\mathbf{R}} /\mathrm{Tr} ( \widehat{\mathbf{R}} )-\mathbf{R}/\mathrm{Tr} \left( \mathbf{R} \right) \right\Vert^2 \right\} }{ \left\Vert \mathbf{R}/\mathrm{Tr} \left( \mathbf{R} \right) \right\Vert^2}. 
    \end{split}
    \label{MMSE}
\end{equation}
Since the true CM of the real data is unknown, we use synthetic  data here. Considering the model in Sec. II, the samples are generated according to $\mathbf{y}_l=\sqrt{\tau_l} \mathbf{u}_l+\mathbf{n}_l, l=1,2,\cdots,L$,
where $\mathbf{u}_l$ is generated by (\ref{xl}) and $\mathbf{n}_l$ denotes the additive white Gaussian noise. Then the corresponding true CM is given by (\ref{Rxpst}).
According to Fig. \ref{fig_fitness}, the sea clutter fits the CG distribution well. Therefore, we assume that the texture $\tau_l$ follows a Gamma distribution \cite{249129} of shape parameter $\nu$ and scale parameter $1/\nu$, i.e., $\tau_l \sim \Gamma (\nu,1/\nu)$, $\mathbf{u}_l \sim \mathcal{CN}(\mathbf{0}, \mathbf{R})$. 
 The generated samples $\{\mathbf{y}_l\}$ follow a zero-mean CES distribution. The estimated sub-CMs $\widehat{\mathbf{R}}_{\mathit{st}}^{(k)} $ and $\widehat{\mathbf{R}}_p^{(k)} $ in (\ref{Ckone}) are initialized as identity matrices for simplicity but other initialization can produce similar results.

Here we set $N_s=1$, $N_t=8$, $N_p=3$. The polarization parameters in (\ref{Rp}) are set as $\rho_c=0.89$, $\gamma_c=0.61$ and $\delta_c=0.16$.  Other radar parameters include the carrier frequency 1.2 GHz,
wavelength 0.25 m,
PRF 2000 Hz, platform velocity 125 m/s and CNR 30 dB. In the rest of this section,  for terminating the iterations, we choose the threshold $\delta$ in (\ref{stop}) as $10^{-3}$ and $K_{\max}=15$. 
For the RSKE, in addition to the KOAS and CV choices of the shrinkage factors, the oracle choice of the shrinkage factors is also considered, which  minimizes the NMSE defined in (\ref{MMSE}) at each iteration under the assumption that the true CM is known.

\begin{table}[!t]
	\renewcommand{\arraystretch}{1.3}
	\caption{Shrinkage coefficient}
	\label{tablesc}
	\centering
	\begin{tabular}{|l||c|c|}
		\hline
		Algorithm  & $\rho_{\textit{st}}$ & $\rho_p$\\
		\hline
		RSKE, KOAS(KNSCM) & 0.0344& 0.2732\\
		RSKE, KOAS(KMLE) & 0.0293& 0.2556\\
		RSKE, CV(KNSCM) & 0.0583& 0.3379\\
		RSKE, CV(KMLE) & 0.0363& 0.3541\\
		RSKE, Oracle & 0& 0.4\\
		\hline
	\end{tabular}
\end{table}

\begin{figure}[h]
	\centering 
	\includegraphics[width=2.5in]{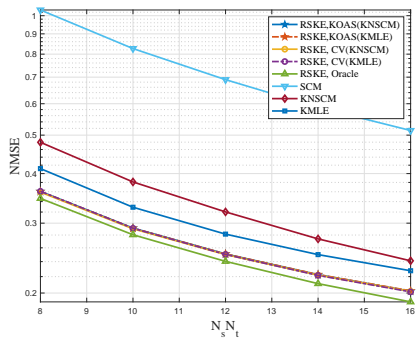}\
	\caption{NMSE versus the space-time number.} 
	\label{fig_Dim}
\end{figure}

Fig. \ref{fig_mse} shows the NMSE performance under different numbers of samples $L$. 
For each abscissa, 2000 Monte-Carlo experiments are performed. 
Note that even a small numerical gap in the NMSE performance may lead to large error between the estimated result and the true CM since the NMSE is normalized.
 We can see that the proposed RSKE can improve the estimation accuracy as compared with several existing estimators in different cases. The CV choices of the shrinkage factors can produce near-oracle performance.  
The performance with $\mathrm{KOAS}$ and $\mathrm{CV}$ depends on the choice of the plug-in estimates used and $\mathrm{CV}$ performs slightly better than $\mathrm{KOAS}$.

\begin{figure}[h]
\centering 
    \includegraphics[width=2.5in]{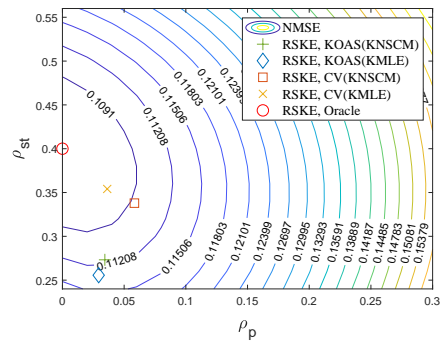}\ 
\caption{NMSE versus $\rho_{\mathit{st}}$ and $\rho_p$.} 
\label{fig_MSE_grid}
\end{figure}

Fig. \ref{fig_Dim} shows the NMSE versus the the dimension $N_s N_t$ of $\mathbf R_{st}$. Here we fix $N_t=2$, $L=\frac{1}{2}N_s N_t$  and vary $N_s$ from 4 to 8. 
As the dimension and the number of samples increase with a constant ratio, the estimation accuracy is also improved.

Fig. \ref{fig_MSE_grid} shows the NMSE versus $\rho_{\mathit{st}}$ and $\rho_p$. 
Here we fix $L=12$ and other parameters are same as Fig. 4.  
100 Monte-Carlo experiments are performed. The average NMSE achieved by RSKE with different $\rho_{\mathit{st}}$ and $\rho_p$ is demonstrated in Fig. \ref{fig_MSE_grid} where the averages of the shrinkage factors chosen by $\mathrm{KOAS}$ and $\mathrm{CV}$ are also marked.  Each line shows the contour of NMSE.
It confirms that the different plug-in estimators used lead to different shrinkage factors.  Moreover, $\mathrm{CV}$ yields solutions closer to the oracle ones compared to  $\mathrm{KOAS}$. The selected shrinkage coefficients are also listed in Table \ref{tablesc}.

\begin{figure}[h]
\centering 
    \includegraphics[width=2.5in]{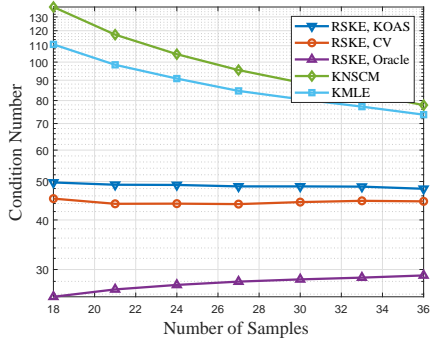}\
\caption{Condition number versus the number of samples.} 
\label{fig_CN}
\end{figure}

Fig. \ref{fig_CN} shows the condition number of the estimated CM of RSKE (with CV, KOAS), KMLE and KNSCM.  
We set the plug-in estimator for $\mathrm{CV}$ and KOAS as KNSCM. 
One can see that the proposed CV and KOAS algorithms yield CM estimates which are better-conditioned than those with  KNSCM and KMLE, especially when the number of samples is small.
As they also improve the NMSE, it is expected that the RSKE with the proposed shrinkage factor choices can improve the performance for applications where the inverse of the CM is required, such as beamforming and spectral estimation applications.

\begin{figure}[h]
	\centering 
	\includegraphics[width=2.5in]{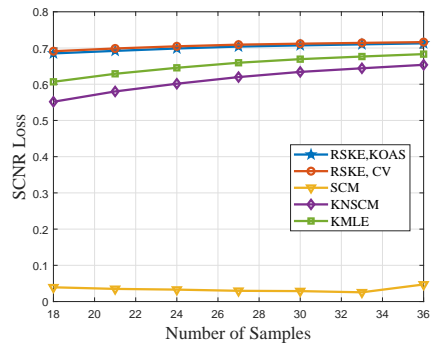}\
	\caption{SCNR loss versus the number of samples.} 
	\label{fig_SL}
\end{figure}

The performance of clutter suppression in PSTAP is often evaluated via 
the normalized SCNR loss \cite{1263229,4101326,9052470} 
\begin{equation}
	\begin{split}
		\mathrm{SCNR}_{\mathrm{loss}}= \frac{\left( \mathbf s ^\mathrm{H} \widehat{\mathbf{R}}^{-1}  \mathbf s   \right)^2 }{\left(\mathbf s ^\mathrm{H}   \widehat{\mathbf{R}}^{-1} \mathbf{R} \widehat{\mathbf{R}}^{-1} \mathbf s  \right)
			\left(\mathbf s ^\mathrm{H}  {\mathbf{R}}^{-1} \mathbf s   \right)}. 
		\label{scnrS1}
	\end{split}
\end{equation}
Clearly its maximum $\mathrm{SCNR}_{\mathrm{loss}} = 1$ is achieved when the covariance matrix is perfectly estimated and a larger value indicates better performance. Parameters are same as those in Fig. 4. For each abscissa, 2000 Monte-Carlo experiments are performed. Fig. \ref{fig_SL} shows the SCNR loss resulted from different covariance estimators. We can see that the proposed RSKE with KOAS and CV can also outperform KNSCM, KMLE and SCM.

\section{Conclusions}

In this paper, we investigate a robust, iterative shrinkage estimator for Kronecker-structured covariance matrices of compound Gaussian data, which is referred to as RSKE. The RSKE can be obtained by minimizing a negative log-likelihood function penalized by Kullback-Leibler divergence and interpreted by integrating linear shrinkage into the fixed-point iterations. 
The conditions for the existence of the RSKE are investigated and the convergence of the iterative solver is investigated. 
We also introduce two methods for choosing the shrinkage factors by exploiting oracle approximating shrinkage (OAS) and cross-validation (CV), respectively. 
The proposed estimators are then applied to polarization radar detection in the real sea clutter context.
Compared with the state-of-the-art estimators, the RSKE achieves better detection performance, more accurate CM estimation and improves the condition number by significantly reducing the number of unknown parameters and integrating shrinkage into the robust estimation.

\appendices
\section{Proof of {Proposition} \ref{ExistenceMain}}
\label{Existence}
In this appendix, we examine the conditions under which a solution to (\ref{fixpoint2}) exists by constructing two auxiliary functions  to lowerbound the  cost function in  (\ref{KMLEProblem0}). 
Let $\lambda_{\mathit{st}}^{(1)} \geq \lambda_{\mathit{st}}^{(2)} \geq \cdots \geq \lambda_{\mathit{st}}^{(N_{\mathit{st}})} $ and  $\lambda_p^{(1)} \geq \lambda_p^{(2)} \geq \cdots \geq \lambda_p^{(N_p)} $ be the eigenvalues of $\widehat{\mathbf{R}}_{\mathit{st}}$ and $\widehat{\mathbf{R}}_p$. Then we have
\begin{equation}
    \begin{split}
&\log  \mathbf y_l^\mathrm{H} \left(\widehat{\mathbf{R}}_{\mathit{st}}\otimes \widehat{\mathbf{R}}_p \right)^{-1} \mathbf y_l \geq \log   \frac{\mathbf y_l^\mathrm{H} \left(\widehat{\mathbf{R}}_{\mathit{st}}\otimes \mathbf{I}_{N_p} \right)^{-1} \mathbf y_l}{\lambda_p^{(1)}} \\
&\geq  \frac{1}{N_p}\sum_{j=1}^{N_p}   \log \mathbf{r}_{j,l}^\mathrm{H}  \widehat{\mathbf{R}}_{\mathit{st}}^{-1}  \mathbf{r}_{j,l}- \log \lambda_p^{(1)}+\log N_p,
    \end{split}
    \label{Inequality1}
\end{equation}
where  we have utilized Jensen's inequality in the last step. 
Similarly, we have 
\begin{equation}
    \begin{split}
&\log  \mathbf y_l^\mathrm{H} \left(\widehat{\mathbf{R}}_{\mathit{st}}\otimes \widehat{\mathbf{R}}_p \right)^{-1} \mathbf y_l\\ 
&\geq  \frac{1}{N_{\mathit{st}}}\sum_{i=1}^{N_{\mathit{st}}}   \log \mathbf{c}_{i,l}^\mathrm{H}  \widehat{\mathbf{R}}_{p}^{-1}  \mathbf{c}_{i,l}- \log \lambda_{\mathit{st}}^{(1)}+\log N_{\mathit{st}}.
    \end{split}
    \label{Inequality2}
\end{equation}
Here we have assumed that none of $\mathbf{r}_{j,l}$ and $\mathbf{c}_{i,l}$ is an all-zero vector, such that 
$\mathbf{r}_{j,l}^\mathrm{H}  \widehat{\mathbf{R}}_{\mathit{st}}^{-1}  \mathbf{r}_{j,l} \ne  0, $ 
$\mathbf{c}_{i,l}^\mathrm{H}  \widehat{\mathbf{R}}_{p}^{-1}  \mathbf{c}_{i,l} \ne 0, \forall i, \forall j, \forall l$. 
Then let us  define the following auxiliary functions: 
\begin{subequations}
\begin{equation}
\nonumber
    \begin{split}
&\mathcal{F}_1 \left(  \widehat{\mathbf{R}}_{\mathit{st}}  \right)\\
& =\frac{N_p L}{2}\log \det \left(  \widehat{\mathbf{R}}_{\mathit{st}}  \right) +  \frac{ \beta_1 N_{\mathit{st}}}{2} \sum_{l=1}^{L}\sum_{j=1}^{N_p}   \log \mathbf{r}_{j,l}^\mathrm{H}  \widehat{\mathbf{R}}_{\mathit{st}}^{-1}  \mathbf{r}_{j,l} \\
&+ \frac{\alpha_{\mathit{st}} L}{2} \mathrm{Tr} \left(  \widehat{\mathbf{R}}_{\mathit{st}}^{-1}  \right) +\frac{\alpha_{\mathit{st}} L}{2} \log \det \left(  \widehat{\mathbf{R}}_{\mathit{st}}  \right) -\frac{\beta_2 L N}{2}  \log \lambda_{\mathit{st}}^{(1)}, \\
    \end{split}
    \label{mathcalF1}
\end{equation} 
\begin{equation}
\nonumber
    \begin{split}
&\mathcal{F}_2 \left(  \widehat{\mathbf{R}}_p  \right)\\
& =\frac{N_{\mathit{st}} L}{2}\log \det \left(  \widehat{\mathbf{R}}_p  \right) +\frac{ \beta_2 N_p}{2} \sum_{l=1}^{L} \sum_{i=1}^{N_{\mathit{st}}}   \log \mathbf{c}_{i,l}^\mathrm{H}  \widehat{\mathbf{R}}_{p}^{-1}  \mathbf{c}_{i,l} \\
&+ \frac{\alpha_p L}{2}\mathrm{Tr} \left(  \widehat{\mathbf{R}}_p^{-1}  \right) +\frac{\alpha_p  L}{2} \log \det \left(  \widehat{\mathbf{R}}_p  \right)- \frac{\beta_1 L N}{2} \log \lambda_p^{(1)}, \\
    \end{split}
    \label{mathcalF2}
\end{equation} 
\end{subequations}
 where $\beta_1+\beta_2=1$ and $\beta_1, \beta_2 \in [0,1]$.  
 From (\ref{Inequality1}) and (\ref{Inequality2}), we have 
\begin{equation}
\nonumber
    \begin{split}
&\mathcal{L}  (  \widehat{\mathbf{R}}_{\mathit{st}}, \widehat{\mathbf{R}}_p   ) \\
&\geq \frac{2}{L} \left(\mathcal{F}_1 (  \widehat{\mathbf{R}}_{\mathit{st}}   )  +\mathcal{F}_2(\widehat{\mathbf{R}}_p) \right)+N(\beta_1 \log N_p+\beta_2 \log N_{\mathit{st}}).
    \end{split}
\end{equation} 
Since  $L$, $N$, $N_{\mathit{st}}$ and $N_p$ are finite, if $\mathcal{F}_1 (  \widehat{\mathbf{R}}_{\mathit{st}} ) \rightarrow +\infty$ and $\mathcal{F}_2(\widehat{\mathbf{R}}_p)\rightarrow +\infty$, then $\mathcal{L}  (  \widehat{\mathbf{R}}_{\mathit{st}}, \widehat{\mathbf{R}}_p) \rightarrow +\infty$.  In the following, we check the conditions under which $\mathcal{F}_1 (  \widehat{\mathbf{R}}_{\mathit{st}} ) \rightarrow +\infty $ and $\mathcal{F}_2(\widehat{\mathbf{R}}_p) \rightarrow +\infty$ on the boundary of the set of positive-definite, Hermitian matrices. 
 Note that $\mathcal{F}_1$ and $\mathcal{F}_2$ are similar to the first equation of \cite[Appendix A]{6879466}.

 Denote the eigenvectors corresponding to $\lambda_{\mathit{st}}^{(i)}$ and $\lambda_p^{(j)}$ by $\mathbf{v}_{\mathit{st}}^{(i)}$ and $\mathbf{v}_p^{(i)}$, respectively, for $\widehat{\mathbf{R}}_{\mathit{st}}$ and $\widehat{\mathbf{R}}_p$. Then denote the subspace spanned by $\{ \mathbf{v}_{\mathit{st}}^{(1)}, \cdots, \mathbf{v}_{\mathit{st}}^{(i)} \}$ and $\{ \mathbf{v}_p^{(1)}, \cdots, \mathbf{v}_p^{(j)} \}$ as $\mathcal{S}_{\mathit{st}}^{(i)}$ and $\mathcal{S}_p^{(j)}$, respectively. Formally, define $\{ r_{\mathit{st}}, s_{\mathit{st}} \}$  with  $1\leq r_{\mathit{st}} \leq s_{\mathit{st}} \leq N_{\mathit{st}}$, such that $\lambda_{\mathit{st}}^{(i)} \to \infty$ for $i \in [1, r_{\mathit{st}}]$, $\lambda_{\mathit{st}}^{(i)}$ is bounded for $i \in (r_{\mathit{st}}, s_{\mathit{st}}]$ and $\lambda_{\mathit{st}}^{(i)} \to 0$ for $i \in (s_{\mathit{st}}, N_{\mathit{st}}]$. Similarly, define $\{r_p, s_p\}$ for $\lambda_p^{(j)}$. Here we consider the case with $r_{\mathit{st}}\geq 1$, i.e., there exists at least one eigenvalue diverging, following  \cite{6879466}, in order to examine the condition for  $\mathcal{F}_1 (  \widehat{\mathbf{R}}_{\mathit{st}} ) \rightarrow +\infty$ at the boundary of feasible set for $\widehat{\mathbf{R}}_{\mathit{st}}$.

 Define 
$
 \mathcal{G}_1 (  \widehat{\mathbf{R}}_{\mathit{st}}  )=\exp (-\mathcal{F}_1 (  \widehat{\mathbf{R}}_{\mathit{st}}  ))
$
 and 
$
 \mathcal{G}_2 (  \widehat{\mathbf{R}}_p  )=\exp (-\mathcal{F}_2 (  \widehat{\mathbf{R}}_p  )).
$

Clearly,  $\mathcal{F}_1 (  \widehat{\mathbf{R}}_{\mathit{st}}  ) \to +\infty$ is equivalent to 
 $\mathcal{G}_1(  \widehat{\mathbf{R}}_{\mathit{st}}  ) \to 0$. From \cite[Appendix A]{6879466}, the condition for $\mathcal{G}_1(  \widehat{\mathbf{R}}_{\mathit{st}}  ) \to 0$ can be checked by examining the infinitesimal equivalence of $\mathcal{G}_1(  \widehat{\mathbf{R}}_{\mathit{st}} )$ in terms of the eigenvalues $\lambda_{\mathit{st}}^{(i)}$ of 
$\widehat{\mathbf{R}}_{\mathit{st}}$. From (36) in \cite[Appendix A]{6879466}, 
 $\mathcal{G}_1(  \widehat{\mathbf{R}}_{\mathit{st}}  ) \to 0$ 
 if the orders of all the eigenvalues $\lambda_{\mathit{st}}^{(i)}   \to \infty$ in the infinitesimal equivalence are negative and those of $\lambda_{\mathit{st}}^{(i)} \to 0$ are positive.
Following this argument, we invoke (36)  in  \cite[Appendix A]{6879466} by letting $N=L N_p$, $K=N_{\mathit{st}}$, $\rho(s)=\frac{\beta_1 N_{\mathit{st}}}{2} \log (s)$, $h_1(s)=s$, $\alpha=\alpha_1=\frac{\alpha_{\mathit{st}} L}{2}$ and $\mathbf{A}_1 =\mathbf{I}_{N_{\mathit{st}}}$, and hence $a_\rho=a_\rho'=\beta_1 N_{\mathit{st}}$ and $a_1=+\infty$, $a_1'=0$\footnote{$(a_\rho, a_\rho')$ and $(a_1, a_1')$ are respectively defined for $\rho(s)$ and $\alpha h_1(s)$ according to \cite[Definition 2]{6879466}}. Note also that for any $\epsilon>0$, 
\[ 
(\lambda_{\mathit{st}}^{(1)} )^{\frac{\beta_2 N L}{2}}=o\left( (\varphi_{\mathit{st}}^{(1)} )^{-\frac{\beta_2 N L}{2} -\epsilon}\right)=o\left( (\varphi_{\mathit{st}}^{(r)} )^{-\frac{\beta_2 N L}{2} -\epsilon}\right),
\]
where $o(\cdot)$ denotes the  higher order infinitesimal and $\varphi_{\mathit{st}}^{(i)}\triangleq  (\lambda_{\mathit{st}}^{(i)} )^{-1}$.  
 Then we impose the same condition as the first line\footnote{The second line of  (36) in \cite[Appendix A]{6879466} is always met since $a_1=+\infty$ in this paper.} of (36)  in \cite[Appendix A]{6879466}, i.e.,
 \begin{equation}
 \nonumber
    \begin{split}
  &\left(\frac{L N_p}{2} +\frac{\alpha_{\mathit{st}} L}{2} -\epsilon \right) d -\frac{\beta_1  N_{\mathit{st}}+\epsilon}{2} L N_p P_{L N_p}(\mathcal{S}_{\mathit{st}}^{(d)})\\
  &-\frac{\beta_2 N L}{2}- \epsilon\geq 0, d=1,\cdots,N_{\mathit{st}}-1.
     \end{split}
\end{equation}
Under this condition,  $\mathcal{G}_1(  \widehat{\mathbf{R}}_{\mathit{st}}  )$ goes to zero , i.e., $\mathcal{F}_1 (  \widehat{\mathbf{R}}_{\mathit{st}}  ) \to +\infty$ on the boundary of positive-definite and Hermitian $\widehat{\mathbf{R}}_{\mathit{st}}$ \cite{6879466}. Letting $\epsilon \to 0$ and rearranging the terms, one has
 \begin{equation}
    \begin{split}
    P_{L N_p}\left(\mathcal{S}_{\mathit{st}}^{(d)} \right) < \frac{ \left(L N_p+\alpha_{\mathit{st}} L \right)d-\beta_2 L N}{\beta_1   L N },
     \end{split}
     \label{conditionAA1}
\end{equation}
for arbitrary $d=1,\cdots,N_{\mathit{st}}-1$. Intuitively, this requires that the samples are evenly spread in the subspace spanned by the eigenvectors of $\widehat{\mathbf{R}}_{\mathit{st}}$. The condition (\ref{conditionAA1}) is then rewritten in a general form as (\ref{conditionAA21}). Similarly, we have (\ref{conditionAB21}).

In summary, we have obtained conditions (\ref{conditionAA21}) and (\ref{conditionAB21}) under which the cost function (\ref{KMLEProblem0}) tends to positive infinity at the boundary of the set of positive definite and Hermitian  matrix. By \cite[Lemma 1]{6879466}, these also give a sufficient condition that a solution to (\ref{fixpoint2}) exists.

\section{Proof of {Proposition} \ref{ConvergenceMain}}
\label{Convergence}
In this Appendix, we prove the convergence of the proposed iteration process, following the methodology of \cite{6298979, 6879466}. 
By the concavity of the logarithm function, one has $\log x \leq \log a+ \frac{x}{a} -1, \forall a>0$.
The equality holds when $x=a$.
Then we have 
\begin{equation}
    \begin{split}
&\log \left[ \mathbf{y}_l^\mathrm{H} \left(\widehat{\mathbf{R}}_{\mathit{st}}\otimes \widehat{\mathbf{R}}_p^{(k)}  \right)^{-1} \mathbf{y}_l \right]   \leq   \frac{  \mathbf{y}_l^\mathrm{H} \left(\widehat{\mathbf{R}}_{\mathit{st}}\otimes \widehat{\mathbf{R}}_p^{(k)}  \right)^{-1} \mathbf{y}_l }{\mathbf{y}_l^\mathrm{H} \left(\widehat{\mathbf{R}}_{\mathit{st}}^{(k)}\otimes \widehat{\mathbf{R}}_p^{(k)}  \right)^{-1} \mathbf{y}_l } \\
 + &\log \left[\mathbf{y}_l^\mathrm{H} \left(\widehat{\mathbf{R}}_{\mathit{st}}^{(k)}\otimes \widehat{\mathbf{R}}_p^{(k)}  \right)^{-1} \mathbf{y}_l  \right]-1, \\
    \end{split}
    \label{NEQ_LOG}
\end{equation} 
where the equality holds when $\widehat{\mathbf{R}}_{\mathit{st}}=\widehat{\mathbf{R}}_{\mathit{st}}^{(k)}$.
We then construct the surrogate function
\begin{equation}
    \begin{split}
&\mathcal{G}_1 \left(  \widehat{\mathbf{R}}_{\mathit{st}} \bigg\vert \widehat{\mathbf{R}}_{\mathit{st}}^{(k)}, \widehat{\mathbf{R}}_p^{(k)}  \right) \\ 
 =&\frac{N_p}{1-\rho_{\mathit{st}}} \log \det \left(  \widehat{\mathbf{R}}_{\mathit{st}}  \right)  +\frac{ N_{\mathit{st}}}{1-\rho_p} \log \det \left(  \widehat{\mathbf{R}}_p^{(k)}  \right)  \\
&+ \frac{N}{L} \sum_{l=1}^{L}  \frac{  \mathbf{y}_l^\mathrm{H} \left(\widehat{\mathbf{R}}_{\mathit{st}}\otimes \widehat{\mathbf{R}}_p^{(k)}  \right)^{-1} \mathbf{y}_l }{\mathbf{y}_l^\mathrm{H} \left(\widehat{\mathbf{R}}_{\mathit{st}}^{(k)}\otimes \widehat{\mathbf{R}}_p^{(k)}  \right)^{-1} \mathbf{y}_l } \\
&+\frac{N}{L} \sum_{l=1}^{L} \log \left[\mathbf{y}_l^\mathrm{H} \left(\widehat{\mathbf{R}}_{\mathit{st}}^{(k)}\otimes \widehat{\mathbf{R}}_p^{(k)}  \right)^{-1} \mathbf{y}_l  \right]-N\\
&+ \frac{N_p \rho_{\mathit{st}}}{1-\rho_{\mathit{st}}} \mathrm{Tr} \left(  \widehat{\mathbf{R}}_{\mathit{st}}^{-1}  \right) + \frac{N_{\mathit{st}} \rho_p}{1-\rho_p}  \mathrm{Tr} \left(  \left( \widehat{\mathbf{R}}_p^{(k)} \right)^{-1}  \right).\\
    \end{split}
    \label{Gak}
\end{equation} 
Recalling (\ref{NEQ_LOG}), we have
\begin{equation}
    \begin{split}
\mathcal{L} \left(  \widehat{\mathbf{R}}_{\mathit{st}}, \widehat{\mathbf{R}}_p^{(k)}  \right)\leq \mathcal{G}_1 \left(  \widehat{\mathbf{R}}_{\mathit{st}}\bigg\vert \widehat{\mathbf{R}}_{\mathit{st}}^{(k)}, \widehat{\mathbf{R}}_p^{(k)}  \right), 
 \end{split}
\end{equation} 
and the equality holds when $\widehat{\mathbf{R}}_{\mathit{st}}=\widehat{\mathbf{R}}_{\mathit{st}}^{(k)}$, i.e.,
\begin{equation}
    \begin{split}
\mathcal{L} \left(  \widehat{\mathbf{R}}_{\mathit{st}}^{(k)}, \widehat{\mathbf{R}}_p^{(k)}  \right)=\mathcal{G}_1 \left(  \widehat{\mathbf{R}}_{\mathit{st}}^{(k)}\bigg\vert \widehat{\mathbf{R}}_{\mathit{st}}^{(k)}, \widehat{\mathbf{R}}_p^{(k)}  \right).
 \end{split}
\end{equation} 
It is easy to verify that the minimizer of (\ref{Gak}) is exactly (\ref{iterA}) by setting the gradient of  (\ref{Gak}) with respect to $\widehat{\mathbf{R}}_{\mathit{st}}$ to zero. It follows that
\begin{equation}
    \begin{split}
\widehat{\mathbf{R}}_{\mathit{st}}^{(k+1)}=\arg \min_{\widehat{\mathbf{R}}_{\mathit{st}}} \mathcal{G}_1 \left(  \widehat{\mathbf{R}}_{\mathit{st}}\bigg\vert \widehat{\mathbf{R}}_{\mathit{st}}^{(k)}, \widehat{\mathbf{R}}_p^{(k)}   \right).
 \end{split}
\end{equation} 
Therefore, 
\begin{equation}
    \begin{split}
&\mathcal{L} \left(  \widehat{\mathbf{R}}_{\mathit{st}}^{(k+1)} , \widehat{\mathbf{R}}_p^{(k)}  \right)\leq \mathcal{G}_1 \left(  \widehat{\mathbf{R}}_{\mathit{st}}^{(k+1)}\bigg\vert \widehat{\mathbf{R}}_{\mathit{st}}^{(k)}, \widehat{\mathbf{R}}_p^{(k)}  \right)\\
&= \min_{\widehat{\mathbf{R}}_{\mathit{st}}} \mathcal{G}_1 \left(  \widehat{\mathbf{R}}_{\mathit{st}}\bigg\vert \widehat{\mathbf{R}}_{\mathit{st}}^{(k)}, \widehat{\mathbf{R}}_p^{(k)}  \right) \leq \mathcal{G}_1 \left(  \widehat{\mathbf{R}}_{\mathit{st}}^{(k)}\bigg\vert \widehat{\mathbf{R}}_{\mathit{st}}^{(k)}, \widehat{\mathbf{R}}_p^{(k)}   \right)\\
&=\mathcal{L} \left(  \widehat{\mathbf{R}}_{\mathit{st}}^{(k)}, \widehat{\mathbf{R}}_p^{(k)}  \right).
 \end{split}
 \label{IterDJ1}
\end{equation} 
Then define
\begin{equation}
    \begin{split}
&\mathcal{G}_2 \left(   \widehat{\mathbf{R}}_p\bigg\vert \widehat{\mathbf{R}}_{\mathit{st}}^{(k+1)}, \widehat{\mathbf{R}}_p^{(k)}   \right) \\
=& \frac{N_p}{1-\rho_{\mathit{st}}} \log \det \left(  \widehat{\mathbf{R}}_{\mathit{st}}^{(k+1)}  \right)
  +\frac{ N_{\mathit{st}}}{1-\rho_p} \log \det \left(  \widehat{\mathbf{R}}_p  \right) 
  \\ 
  & + \frac{N}{L} \sum_{l=1}^{L}  \frac{  \mathbf{y}_l^\mathrm{H} \left(\widehat{\mathbf{R}}_{\mathit{st}}^{(k+1)}\otimes \widehat{\mathbf{R}}_p  \right)^{-1} \mathbf{y}_l }{\mathbf{y}_l^\mathrm{H} \left(\widehat{\mathbf{R}}_{\mathit{st}}^{(k+1)}\otimes \widehat{\mathbf{R}}_p^{(k)}  \right)^{-1} \mathbf{y}_l } \\
&+\frac{N}{L} \sum_{l=1}^{L} \log \left[\mathbf{y}_l^\mathrm{H} \left(\widehat{\mathbf{R}}_{\mathit{st}}^{(k+1)}\otimes \widehat{\mathbf{R}}_p^{(k)}  \right)^{-1} \mathbf{y}_l  \right]-N\\
&+ \frac{N_p \rho_{\mathit{st}}}{1-\rho_{\mathit{st}}}  \mathrm{Tr} \left( \left( \widehat{\mathbf{R}}_{\mathit{st}}^{(k+1)} \right)^{-1}  \right) + \frac{N_{\mathit{st}} \rho_p}{1-\rho_p}  \mathrm{Tr} \left(  \widehat{\mathbf{R}}_p^{-1}  \right).\\
    \end{split}
    \label{Gbk}
\end{equation} 
Similarly,  we can verify that the minimizer of (\ref{Gbk}) is exactly (\ref{iterB}), and 
\begin{equation}
    \begin{split}
\mathcal{L} \left(  \widehat{\mathbf{R}}_{\mathit{st}}^{(k+1)}, \widehat{\mathbf{R}}_p  \right)\leq \mathcal{G}_2  \left(  \widehat{\mathbf{R}}_p\bigg\vert \widehat{\mathbf{R}}_{\mathit{st}}^{(k+1)}, \widehat{\mathbf{R}}_p^{(k)}  \right), 
 \end{split}
\end{equation} 
where the equality 
holds when $\widehat{\mathbf{R}}_p=\widehat{\mathbf{R}}_p^{(k)}$, i.e.,
\begin{equation}
    \begin{split}
\mathcal{L} \left(  \widehat{\mathbf{R}}_{\mathit{st}}^{(k+1)}, \widehat{\mathbf{R}}_p^{(k)}  \right)=\mathcal{G}_2  \left(  \widehat{\mathbf{R}}_p^{(k)}\bigg\vert \widehat{\mathbf{R}}_{\mathit{st}}^{(k+1)}, \widehat{\mathbf{R}}_p^{(k)}  \right).
 \end{split}
\end{equation} 
 It follows that
\begin{equation}
    \begin{split}
&\mathcal{L} \left(  \widehat{\mathbf{R}}_{\mathit{st}}^{(k+1)} , \widehat{\mathbf{R}}_p^{(k+1)}  \right)\leq \mathcal{G}_2  \left(  \widehat{\mathbf{R}}_p^{(k+1)}\bigg\vert \widehat{\mathbf{R}}_{\mathit{st}}^{(k+1)}, \widehat{\mathbf{R}}_p^{(k)}  \right)\\
&= \min_{\widehat{\mathbf{R}}_p} \mathcal{G}_2  \left(  \widehat{\mathbf{R}}_p\bigg\vert \widehat{\mathbf{R}}_{\mathit{st}}^{(k+1)}, \widehat{\mathbf{R}}_p^{(k)}  \right)\leq \mathcal{G}_2  \left(  \widehat{\mathbf{R}}_p^{(k)}\bigg\vert \widehat{\mathbf{R}}_{\mathit{st}}^{(k+1)}, \widehat{\mathbf{R}}_p^{(k)}   \right)\\
&=\mathcal{L} \left(  \widehat{\mathbf{R}}_{\mathit{st}}^{(k+1)}, \widehat{\mathbf{R}}_p^{(k)}  \right).
 \end{split}
 \label{IterDJ2}
\end{equation} 
Combining (\ref{IterDJ1}) and (\ref{IterDJ2}), we have
\begin{equation}
    \begin{split}
\mathcal{L} \left(  \widehat{\mathbf{R}}_{\mathit{st}}^{(k+1)} , \widehat{\mathbf{R}}_p^{(k+1)}  \right) \leq  \mathcal{L} \left(  \widehat{\mathbf{R}}_{\mathit{st}}^{(k)} , \widehat{\mathbf{R}}_p^{(k)}  \right), 
 \end{split}
\end{equation} 
i.e., the penalized log-likelihood function $\mathcal{L} (  \widehat{\mathbf{R}}_{\mathit{st}}, \widehat{\mathbf{R}}_p  ) $ in (\ref{KMLEProblem0}) is decreasing with iterations.

Since $\mathcal{L} (  \widehat{\mathbf{R}}_{\mathit{st}}, \widehat{\mathbf{R}}_p  ) $ is g-convex, its minimizer exists and denote it by $(\widehat{\mathbf{R}}_{\mathit{st}}^{\infty}, \widehat{\mathbf{R}}_p^{\infty})$. 
Then $\mathcal{L}(\widehat{\mathbf{R}}_{\mathit{st}}^{\infty}, \widehat{\mathbf{R}}_p^{\infty})$  lower bounds the sequence $\{\mathcal{L} (\widehat{\mathbf{R}}_{\mathit{st}}^{(k)}, \widehat{\mathbf{R}}_p^{(k)}), k=1,2,\cdots\}$. This indicates that the decreasing sequence  $\{  \mathcal{L} (\widehat{\mathbf{R}}_{\mathit{st}}^{(k)}, \widehat{\mathbf{R}}_p^{(k)}) \}  $  is   bounded by an infimum. Then according to the monotone convergence theorem \cite{bibby_1974}, the sequence will converge to the infimum as $k$ increases, i.e.,  $(\widehat{\mathbf{R}}_{\mathit{st}}^{(k)}, \widehat{\mathbf{R}}_p^{(k)})$ will converge to the minimizer of $\mathcal{L} (  \widehat{\mathbf{R}}_{\mathit{st}}, \widehat{\mathbf{R}}_p  )$, i.e., the solution to (\ref{fixpoint2}).

\section{Proof of {Proposition} \ref{ProKOAS}}
\label{proofPro1}
We here complete the proof by exploiting results from random matrix theory. Following \cite{LEDOIT2003603}, when the true covariance matrix $\mathbf R_{\mathit{st}}$ and $\mathbf R_p$ are known, the oracle shrinkage factor $\rho_p^\star$, i.e., the solution to (\ref{rhoBequation}), is given by
\begin{equation}
    \begin{split}
\rho_p^{\star}&=\frac{\mathbb{E} \left\{ \mathrm{Re}\left( \mathrm{Tr} \left(  \left( \mathbf{I}_{N_p}-\mathbf{C}_p \right)  \left({\mathbf{R}}_p-\mathbf{C}_p \right)^\mathrm{H} \right) \right) \right\}}{\mathbb{E} \left\{  \Vert \mathbf{I}_{N_p}-  \mathbf{C}_p    \Vert^2 \right\}}\\
&=\frac{E_1
-E_2
-E_3
+\mathrm{Tr}\left(   {\mathbf{R}}_p    \right)
}{E_1
-2E_2+N_p
}, 
    \end{split}
    \label{rhosp}
\end{equation} 
where $\mathrm{Re}(\cdot)$ denotes the real part and 
\begin{equation}
    \begin{split}
    	&E_{1}=\mathbb{E} \left\{   \mathrm{Tr}\left(   \mathbf{C}_p^2    \right)    \right\},E_2=\mathbb{E} \left\{  \mathrm{Re}\left(  \mathrm{Tr}\left(   \mathbf{C}_p    \right)  \right)  \right\},\\
    	&E_3=\mathbb{E} \left\{  \mathrm{Re}\left( \mathrm{Tr}\left(   \mathbf{C}_p{\mathbf{R}}_p^\mathrm{H}     \right)  \right)\right\}
    \end{split}
    \label{ApE1}
\end{equation} 
and $\mathbf{C}_p$ is defined by (\ref{CB}). The resulting optimal shrinkage estimate can be interpreted as the projection of the true CM onto the linear space spanned by $\mathbf{C}_p$ and $\mathbf{I}_{N_p}$.

Let the eigen-decomposition of $\mathbf{R}$, $\mathbf{R}_{\mathit{st}}$ and $\mathbf{R}_p$ be $\mathbf{R}=\mathbf{V} \mathbf{\Lambda} \mathbf{V}^\mathrm{H}$, $\mathbf{R}_{\mathit{st}}=\mathbf{V}_{\mathit{st}} \mathbf{\Lambda}_{\mathit{st}} \mathbf{V}_{\mathit{st}}^\mathrm{H}$, and $\mathbf{R}_p=\mathbf{V}_p \mathbf{\Lambda}_p \mathbf{V}_p^\mathrm{H}$, respectively. 
Then, we define $\mathbf{z}_l=\frac{\mathbf{D}^{-1}\mathbf{y}_l}{   \left\Vert \mathbf{D}^{-1}\mathbf{y}_l   \right\Vert_2}$,
where  $\mathbf{D}=\mathbf{V}\mathbf{\Lambda}^{\frac{1}{2}}$.
It is easy to see that $\Vert \mathbf{z}_l \Vert_2=1$ and $\{ \mathbf{z}_l \}$ are independent of each other. Moreover, the whiten vectors $\{ \mathbf{z}_l \}$ are isotropically distributed \cite{746779} and satisfy \cite{5484583,5743027} 
\begin{equation}
    \begin{split}
    &\mathbb{E}\left\{  \mathbf{z}_l \mathbf{z}_l^\mathrm{H} \right\}=\frac{1}{N} \mathbf{I}_{N},\\
     &\mathbb{E}\left\{  \left( \mathbf{z}_l^\mathrm{H} \mathbf{\Lambda} \mathbf{z}_l \right) ^2\right\}=\frac{  \mathrm{Tr} \left(      \mathbf{R}^2 
               \right) +   \mathrm{Tr}^2 \left(        \mathbf{R}      \right)}{N(N+1)},\\
     &\mathbb{E}\left\{  \left( \mathbf{z}_{l}^\mathrm{H} \mathbf{\Lambda}  \mathbf{z}_{q} \right) ^2\right\}=\frac{1}{N^2} \mathrm{Tr}\left(        \mathbf{R}^2         \right) , l \neq q. 
    \end{split}
    \label{Rzlzl}
\end{equation} 
Note that $\mathbf{D}=\mathbf{D}_{\mathit{st}} \otimes \mathbf{D}_{p}$,
where $\mathbf{D}_{\mathit{st}}={\mathbf{V}}_{\mathit{st}} {\mathbf{\Lambda}}_{\mathit{st}}^{\frac{1}{2}}$, $\mathbf{D}_p={\mathbf{V}}_p {\mathbf{\Lambda}}_p^{\frac{1}{2}}$.
We then reshape $\mathbf{z}_l$ into a matrix satisfying 
\begin{equation}
    \begin{split}
\mathbf{Z}_l &=\mathrm{unvec}_{N_p N_{\mathit{st}}}(\mathbf{z}_l)
=\frac{ \mathbf{D}_p^{-1} \mathbf{Y}_l \left(\mathbf{D}_{\mathit{st}}^{-1}   \right)^\mathrm{H}}{\left\Vert \mathbf{D}^{-1}\mathbf{y}_l   \right\Vert_2 },
    \end{split}
\label{Zl}
\end{equation} 
which can be easily verified by vectorizing both sides of (\ref{Zl}).

In order to determine the shrinkage factor for the robust shrinkage estimator of unstructured CM, \cite{5743027} analyzed the feature of $\mathbf{Z}_l$ where it reduces to a vector. We here extend the analysis to the more general case of matrix-valued $\mathbf{Z}_l $ by exploiting random matrix theory and properties of Kronecker product.
Let ${z}_l^{(i)}$ be the $i$th entry of $\mathbf{z}_l$. 
From (\ref{Rzlzl}), one has
\begin{equation}
    \begin{split}
    \mathbb{E}  \left\{{z}_l^{(i)} \left({z}_l^{(j)} \right)^{*} \right\}=\left\{\begin{matrix}
1/N & i=j\\
0 & i\neq j\\
\end{matrix}\right.  .
    \end{split}
\end{equation} 
This indicates that $\{{z}_l^{(i)} \}_{i=1}^N$ are i.i.d. 
with zero mean and variance $1/N$. Consequently, 
we have 
\begin{equation}
    \begin{split}
\mathbb{E}  \left\{  \mathbf{Z}_l   \mathbf{Z}_l^\mathrm{H} \right\}=\frac{N_{\mathit{st}}}{N} \mathbf{I}_{N_p}, \mathbb{E}  \left\{  \mathbf{Z}_l^\mathrm{H}    \mathbf{Z}_l\right\}=\frac{N_p}{N} \mathbf{I}_{N_{\mathit{st}}}.
    \end{split}
    \label{exppppp}
\end{equation} 
Note that $ \left\Vert \mathbf{D}^{-1}\mathbf{y}_l   \right\Vert_2^2=\mathbf{y}_l^\mathrm{H} \left( {\mathbf{R}}_{\mathit{st}} \otimes {\mathbf{R}}_p \right)^{-1}  \mathbf{y}_l$,
and  we have
\begin{equation}
    \begin{split}
		&\frac{\mathbf{Y}_l {\mathbf{R}}_{\mathit{st}}^{-1} \mathbf{Y}_l^\mathrm{H}}{\mathbf{y}_l^\mathrm{H} \left( {\mathbf{R}}_{\mathit{st}} \otimes {\mathbf{R}}_p \right)^{-1}  \mathbf{y}_l}=\mathbf{D}_p \mathbf{Z}_{l} \mathbf{Z}_{l}^\mathrm{H}  \mathbf{D}_p^\mathrm{H},\\
		&\frac{\mathbf{Y}_l^\mathrm{H} {\mathbf{R}}_p^{-1} \mathbf{Y}_l}{\mathbf{y}_l^\mathrm{H} \left( {\mathbf{R}}_{\mathit{st}} \otimes {\mathbf{R}}_p \right)^{-1}  \mathbf{y}_l}=\mathbf{D}_{\mathit{st}} \mathbf{Z}_{l}^\mathrm{H} \mathbf{Z}_{l}  \mathbf{D}_{\mathit{st}}^\mathrm{H},
    \end{split}
    \label{YYZZ}
\end{equation}

Note that $\mathbb{E}\{\cdot\}$, $\mathrm{Re}(\cdot)$  and $\mathrm{Tr}(\cdot)$  are exchangeable to each other. Substituting (\ref{YYZZ}) into  (\ref{ApE1}), one has
\begin{equation}
    \begin{split}
		E_2&=\mathrm{Tr}  \left( \frac{N}{ L N_{\mathit{st}}} \sum_{l=1}^L \mathbf{D}_p \mathbb{E}   \left( \mathbf{Z}_{l} \mathbf{Z}_{l}^\mathrm{H}  \right) \mathbf{D}_p^\mathrm{H}        \right) =\mathrm{Tr}\left(    {\mathbf{R}}_p    \right),\\
		E_3&=\mathrm{Tr}\left(    {\mathbf{R}}_p^2 \right),
    \end{split}
    \label{AppendixE2}
\end{equation} 
From \cite{Hiai00asymptoticfreeness,tulino2004random}, we have 
\begin{equation}
    \begin{split}
\mathbb{E}\left\{    \left| z_l^{(i)} \right|^4   \right\}=\frac{2}{N(N+1)},\mathbb{E}\left\{    \left| z_l^{(i)} \right|^2    \left| z_l^{(j)} \right|^2    \right\}=\frac{1}{N(N+1)}. 
    \end{split}
    \label{ApC1}
\end{equation} 
Since $\left\{\mathbf{z}_l \right\}_{l=1}^L$ are i.i.d, we have 
\begin{equation}
    \begin{split}
\mathbb{E}\left\{    \left| z_l^{(i)} \right|^2  \left| z_q^{(i)} \right|^2  \right\}=\frac{1}{N^2},\mathbb{E}\left\{   z_q^{(i)} \left( z_l^{(i)} \right)^*  z_l^{(j)}   \left( z_q^{(j)} \right)^*   \right\}=0. 
    \end{split}
    \label{ApC}
\end{equation} 
Therefore, (\ref{ApE1}) can be rewritten as
\begin{equation}
    \begin{split}
    	&E_{1}=\mathbb{E} \left\{   \mathrm{Tr}\left(   \mathbf{C}_p^2  \right)    \right\}\\
    	&=\left(\frac{N}{ L N_{\mathit{st}}}  \right)^2   \mathbb{E} \left\{    \mathrm{Tr}     \left(    \sum_{l=1}^L \sum_{q=1}^L \mathbf{D}_p \mathbf{Z}_{l} \mathbf{Z}_{l}^\mathrm{H}  \mathbf{D}_p^\mathrm{H}   \mathbf{D}_p \mathbf{Z}_{q} \mathbf{Z}_{q}^\mathrm{H}  \mathbf{D}_p^\mathrm{H}    \right)   \right\}\\
    	&=\left(\frac{N}{ L N_{\mathit{st}}}  \right)^2   \mathbb{E} \left\{   \sum_{l=1}^L \sum_{q=1}^L  \mathrm{Tr}     \left(     \mathbf{Z}_{l}^\mathrm{H}  {\mathbf{\Lambda}}_p  \mathbf{Z}_{q}  \mathbf{Z}_{q}^\mathrm{H}  {\mathbf{\Lambda}}_p  \mathbf{Z}_{l}     \right)   \right\}.
    \end{split}
    \label{AppendixE1}
\end{equation} 
Utilizing \cite[Lemma 1.1]{Hiai00asymptoticfreeness} and substituting (\ref{ApC1}), (\ref{ApC})   into  (\ref{AppendixE1}), $E_1$ is obtained as  (\ref{largeE1})  in the following page. 
\begin{figure*}[!t]
\normalsize
\begin{equation}
    \begin{split}
    	E_{1}
    	&=\left(\frac{N}{ L N_{\mathit{st}}}  \right)^2   \mathbb{E} \left\{   \sum_{l=1}^L \sum_{q=1}^L  \sum_{i=1}^{N_{\mathit{st}}} \sum_{k=1}^{N_{\mathit{st}}}  \sum_{m=1}^{N_p} \sum_{n=1}^{N_p} \left( {\lambda}_p^{(m)} {\lambda}_p^{(n)}    z_q^{({N_p}(k-1)+m)} \left( z_l^{({N_p}(i-1)+m)}  \right)^* z_l^{({N_p}(i-1)+n)} \left( z_q^{({N_p}(k-1)+n)}  \right)^*  \right)  \right\}\\
    	&=\left(\frac{N}{ L N_{\mathit{st}}}  \right)^2   \left[
    	\left(  \frac{2N_{\mathit{st}}L}{N(N+1)}    +    \frac{N_{\mathit{st}}L(L-1)}{N^2}   +    \frac{N_{\mathit{st}}(N_{\mathit{st}}-1)L}{N(N+1)}  +   \frac{N_{\mathit{st}} \left(N_{\mathit{st}}-1\right)L(L-1)}{N^2}   \right)  
    	\left( \sum_{m=1}^{N_p}  \left( {\lambda}_p^{(m)}\right)^2  \right)  \right.\\
    	&+\left.     \left(  \frac{N_{\mathit{st}}L}{N(N+1)}   \right)  
    	 \left(\sum_{m \neq n}  {\lambda}_p^{(m)} {\lambda}_p^{(n)}    \right)  \right] =  	\left( 1- \frac{1}{L(N+1)}   \right)
    	\mathrm{Tr}\left(  {\mathbf{R}}_p^2 \right) 
    	+   \left(    \frac{N}{N_{\mathit{st}}L(N+1)}  \right)  
    	 \mathrm{Tr}^2\left(  {\mathbf{R}}_p \right).\\
    \end{split}
    \label{largeE1}
\end{equation} 
\hrulefill
\vspace*{4pt}
\end{figure*}
Substituting (\ref{largeE1}) and (\ref{AppendixE2}) into  (\ref{rhosp}),  (\ref{rhoBopt0}) is obtained.
Similarly, we can have the optimal $\rho_{\mathit{st}}^\star$, i.e., (\ref{rhoAopt0}).
 The resulting expressions of $\rho_{\mathit{st}}^\star$ and $\rho_p^\star$ can be used to produce the KOAS choice $\rho_{\textit{st}, \mathrm{KOAS}}$ and $\rho_{p, \mathrm{KOAS}}$ by plugging estimates of $\mathbf R_{\mathit{st}}$ and $\mathbf R_p$ into (\ref{rhoopt}).

\section{Proof of {Proposition} \ref{ExpectMain}}
\label{Expect}
This proposition can be proven by combining the results in Appendix \ref{proofPro1}.
Recalling (\ref{expectSAB}), (\ref{exppppp}) and (\ref{YYZZ}), we have
\begin{equation}
    \begin{split}
\mathbb{E}  \left(     \mathbf{S}_{\mathit{st}}      \right) &=N_{\mathit{st}} \mathbf{D}_{\mathit{st}} \mathbb{E}  \left(   \mathbf{Z}_{l}^\mathrm{H} \mathbf{Z}_{l}  \right) \mathbf{D}_{\mathit{st}}^\mathrm{H}=\mathbf{R}_{\mathit{st}},\\
 \mathbb{E}  \left(     \mathbf{S}_p \right) &=N_p \mathbf{D}_p \mathbb{E}  \left(   \mathbf{Z}_{l} \mathbf{Z}_{l}^\mathrm{H}  \right) \mathbf{D}_p^\mathrm{H}=\mathbf{R}_p.
\end{split}
\end{equation} 
Moreover, (\ref{expectcostAB}) can be rewritten as
\begin{subequations}
\label{expectcostAB2}
\begin{equation}
    \begin{split}
\mathcal{J}_{\mathit{st}}\left( \mathbf{\Sigma}_{\mathit{st}}\right)=\mathrm{Tr} \left( \mathbf{\Sigma}_{\mathit{st}}^2 -2\mathrm{Re}\left(\mathbf{\Sigma}_{\mathit{st}}  \mathbb{E} \left( \mathbf{S}_{\mathit{st}}\right) \right) +\mathbb{E} \left( \mathbf{S}_{\mathit{st}}^2 \right)  \right),
    \end{split}
    \label{expectcostA2}
\end{equation} 
\begin{equation}
    \begin{split}
\mathcal{J}_{p}\left( \mathbf{\Sigma}_p\right)=\mathrm{Tr} \left( \mathbf{\Sigma}_p^2 -2 \mathrm{Re}\left( \mathbf{\Sigma}_p  \mathbb{E} \left( \mathbf{S}_p \right)  \right)   +\mathbb{E} \left( \mathbf{S}_p^2 \right)  \right).
    \end{split}
    \label{expectcostB2}
\end{equation} 
\end{subequations}
By setting the derivative of (\ref{expectcostA2}) and (\ref{expectcostB2}) with respect to $\mathbf{\Sigma}_{\mathit{st}} $ and $\mathbf{\Sigma}_p $ to zero, we have the minimizer of (\ref{expectcostAB}) as $\mathbf{\Sigma}_{\mathit{st}}=\mathbf{R}_{\mathit{st}}$ and $\mathbf{\Sigma}_p=\mathbf{R}_p$.



\begin{thebibliography}{10}
	\providecommand{\url}[1]{#1}
	\csname url@samestyle\endcsname
	\providecommand{\newblock}{\relax}
	\providecommand{\bibinfo}[2]{#2}
	\providecommand{\BIBentrySTDinterwordspacing}{\spaceskip=0pt\relax}
	\providecommand{\BIBentryALTinterwordstretchfactor}{4}
	\providecommand{\BIBentryALTinterwordspacing}{\spaceskip=\fontdimen2\font plus
		\BIBentryALTinterwordstretchfactor\fontdimen3\font minus
		\fontdimen4\font\relax}
	\providecommand{\BIBforeignlanguage}[2]{{%
			\expandafter\ifx\csname l@#1\endcsname\relax
			\typeout{** WARNING: IEEEtran.bst: No hyphenation pattern has been}%
			\typeout{** loaded for the language `#1'. Using the pattern for}%
			\typeout{** the default language instead.}%
			\else
			\language=\csname l@#1\endcsname
			\fi
			#2}}
	\providecommand{\BIBdecl}{\relax}
	\BIBdecl
	
	\bibitem{9279243}
	P.~Huang, Z.~Zou, X.-G. Xia, X.~Liu, G.~Liao, and Z.~Xin, ``Multichannel sea
	clutter modeling for spaceborne early warning radar and clutter suppression
	performance analysis,'' \emph{IEEE Transactions on Geoscience and Remote
		Sensing}, pp. 1--18, 2020.
	
	\bibitem{8370234}
	J.~Yin, C.~Unal, M.~Schleiss, and H.~Russchenberg, ``Radar target and moving
	clutter separation based on the low-rank matrix optimization,'' \emph{IEEE
		Transactions on Geoscience and Remote Sensing}, vol.~56, no.~8, pp.
	4765--4780, 2018.
	
	\bibitem{8752101}
	S.~Allabakash, S.~Lim, P.~Yasodha, H.~Kim, and G.~Lee, ``Intermittent clutter
	suppression method based on adaptive harmonic wavelet transform for l-band
	radar wind profiler,'' \emph{IEEE Transactions on Geoscience and Remote
		Sensing}, vol.~57, no.~11, pp. 8546--8556, 2019.
	
	\bibitem{7180351}
	E.~Makhoul, C.~López-Martínez, and A.~Broquetas, ``Exploiting polarimetric
	terrasar-x data for sea clutter characterization,'' \emph{IEEE Transactions
		on Geoscience and Remote Sensing}, vol.~54, no.~1, pp. 358--372, 2016.
	
	\bibitem{5325631}
	J.~Carretero-Moya, J.~Gismero-Menoyo, A.~Blanco-del Campo, and
	A.~Asensio-Lopez, ``Statistical analysis of a high-resolution sea-clutter
	database,'' \emph{IEEE Transactions on Geoscience and Remote Sensing},
	vol.~48, no.~4, pp. 2024--2037, 2010.
	
	\bibitem{zhang2019ship}
	T.~Zhang, L.~Jiang, D.~Xiang, Y.~Ban, L.~Pei, and H.~Xiong, ``Ship detection
	from polsar imagery using the ambiguity removal polarimetric notch filter,''
	\emph{ISPRS Journal of Photogrammetry and Remote Sensing}, vol. 157, pp.
	41--58, 2019.
	
	\bibitem{9085903}
	T.~Zhang, Z.~Yang, H.~Gan, D.~Xiang, S.~Zhu, and J.~Yang, ``Polsar ship
	detection using the joint polarimetric information,'' \emph{IEEE Transactions
		on Geoscience and Remote Sensing}, vol.~58, no.~11, pp. 8225--8241, 2020.
	
	\bibitem{7522075}
	Z.~Xin, G.~Liao, Z.~Yang, Y.~Zhang, and H.~Dang, ``A deterministic sea-clutter
	space–time model based on physical sea surface,'' \emph{IEEE Transactions
		on Geoscience and Remote Sensing}, vol.~54, no.~11, pp. 6659--6673, 2016.
	
	\bibitem{8506468}
	Y.~Yang, S.-P. Xiao, and X.-S. Wang, ``Radar detection of small target in sea
	clutter using orthogonal projection,'' \emph{IEEE Geoscience and Remote
		Sensing Letters}, vol.~16, no.~3, pp. 382--386, 2019.
	
	\bibitem{7126972}
	H.~Ding, J.~Guan, N.~Liu, and G.~Wang, ``New spatial correlation models for sea
	clutter,'' \emph{IEEE Geoscience and Remote Sensing Letters}, vol.~12, no.~9,
	pp. 1833--1837, 2015.
	
	\bibitem{1610833}
	H.~Melief, H.~Greidanus, P.~van Genderen, and P.~Hoogeboom, ``Analysis of sea
	spikes in radar sea clutter data,'' \emph{IEEE Transactions on Geoscience and
		Remote Sensing}, vol.~44, no.~4, pp. 985--993, 2006.
	
	\bibitem{9420730}
	A.~R. Monteith and L.~M.~H. Ulander, ``A tower-based radar study of temporal
	coherence of a boreal forest at p-, l-, and c-bands and linear cross
	polarization,'' \emph{IEEE Transactions on Geoscience and Remote Sensing},
	pp. 1--15, 2021.
	
	\bibitem{9367301}
	N.~Longépé, A.~A. Mouche, L.~Ferro-Famil, and R.~Husson,
	``Co-cross-polarization coherence over the sea surface from sentinel-1 sar
	data: Perspectives for mission calibration and wind field retrieval,''
	\emph{IEEE Transactions on Geoscience and Remote Sensing}, pp. 1--16, 2021.
	
	\bibitem{1661791}
	Y.~Wang and V.~Chandrasekar, ``Polarization isolation requirements for linear
	dual-polarization weather radar in simultaneous transmission mode of
	operation,'' \emph{IEEE Transactions on Geoscience and Remote Sensing},
	vol.~44, no.~8, pp. 2019--2028, 2006.
	
	\bibitem{7150390}
	C.~Lukashin, Z.~Jin, G.~Kopp, D.~G. MacDonnell, and K.~Thome, ``Clarreo
	reflected solar spectrometer: Restrictions for instrument sensitivity to
	polarization,'' \emph{IEEE Transactions on Geoscience and Remote Sensing},
	vol.~53, no.~12, pp. 6703--6709, 2015.
	
	\bibitem{6472767}
	M.~Galletti, D.~Huang, and P.~Kollias, ``Zenith/nadir pointing mm-wave radars:
	Linear or circular polarization?'' \emph{IEEE Transactions on Geoscience and
		Remote Sensing}, vol.~52, no.~1, pp. 628--639, 2014.
	
	\bibitem{1292147}
	A.~D. Maio, G.~Alfano, and E.~Conte, ``Polarization diversity detection in
	compound-{G}aussian clutter,'' \emph{IEEE Transactions on Aerospace and
		Electronic Systems}, vol.~40, no.~1, pp. 114--131, Jan 2004.
	
	\bibitem{XIE2020107401}
	L.~Xie, Z.~He, J.~Tong, J.~Li, and H.~Li, ``Transmitter polarization
	optimization for space-time adaptive processing with diversely polarized
	antenna array,'' \emph{Signal Processing}, vol. 169, p. 107401, 2020.
	
	\bibitem{628782}
	G.~{Noriega} and S.~{Pasupathy}, ``Adaptive estimation of noise covariance
	matrices in real-time preprocessing of geophysical data,'' \emph{IEEE
		Transactions on Geoscience and Remote Sensing}, vol.~35, no.~5, pp.
	1146--1159, 1997.
	
	\bibitem{tadjudin1999covariance}
	S.~Tadjudin and D.~A. Landgrebe, ``Covariance estimation with limited training
	samples,'' \emph{IEEE Transactions on Geoscience and Remote Sensing},
	vol.~37, no.~4, pp. 2113--2118, 1999.
	
	\bibitem{4101326}
	I.~S. Reed, J.~D. Mallett, and L.~E. Brennan, ``{R}apid convergence rate in
	adaptive arrays,'' \emph{IEEE Transactions on Aerospace and Electronic
		Systems}, vol.~10, no.~6, pp. 853--863, Nov 1974.
	
	\bibitem{bickel2008regularized}
	P.~J. Bickel, E.~Levina \emph{et~al.}, ``Regularized estimation of large
	covariance matrices,'' \emph{The Annals of Statistics}, vol.~36, no.~1, pp.
	199--227, 2008.
	
	\bibitem{CHEN199910}
	P.~Chen, W.~L. Melvin, and M.~C. Wicks, ``Screening among multivariate normal
	data,'' \emph{Journal of Multivariate Analysis}, vol.~69, no.~1, pp. 10--29,
	1999.
	
	\bibitem{4104190}
	E.~J. {Kelly}, ``An adaptive detection algorithm,'' \emph{IEEE Transactions on
		Aerospace and Electronic Systems}, vol.~22, no.~2, pp. 115--127, March 1986.
	
	\bibitem{6263313}
	E.~Ollila, D.~E. Tyler, V.~Koivunen, and H.~V. Poor, ``{C}omplex elliptically
	symmetric distributions: Survey, new results and applications,'' \emph{IEEE
		Transactions on Signal Processing}, vol.~60, no.~11, pp. 5597--5625, Nov
	2012.
	
	\bibitem{89019}
	C.~J. {Baker}, ``K-distributed coherent sea clutter,'' \emph{IEE Proceedings F
		- Radar and Signal Processing}, vol. 138, no.~2, pp. 89--92, 1991.
	
	\bibitem{4647137}
	E.~{Conte} and M.~{Longo}, ``Characterisation of radar clutter as a spherically
	invariant random process,'' \emph{IEE Proceedings F - Communications, Radar
		and Signal Processing}, vol. 134, no.~2, pp. 191--197, April 1987.
	
	\bibitem{766928}
	K.~J. {Sangston}, F.~{Gini}, M.~V. {Greco}, and A.~{Farina}, ``Structures for
	radar detection in compound gaussian clutter,'' \emph{IEEE Transactions on
		Aerospace and Electronic Systems}, vol.~35, no.~2, pp. 445--458, 1999.
	
	\bibitem{766939}
	J.~B. {Billingsley}, A.~{Farina}, F.~{Gini}, M.~V. {Greco}, and
	L.~{Verrazzani}, ``Statistical analyses of measured radar ground clutter
	data,'' \emph{IEEE Transactions on Aerospace and Electronic Systems},
	vol.~35, no.~2, pp. 579--593, April 1999.
	
	\bibitem{wu2014training}
	Y.~Wu, T.~Wang, J.~Wu, and J.~Duan, ``Training sample selection for space-time
	adaptive processing in heterogeneous environments,'' \emph{IEEE Geoscience
		and Remote Sensing Letters}, vol.~12, no.~4, pp. 691--695, 2014.
	
	\bibitem{6825699}
	A.~{Aubry}, A.~D. {Maio}, L.~{Pallotta}, and A.~{Farina}, ``Median matrices and
	their application to radar training data selection,'' \emph{IET Radar, Sonar
		Navigation}, vol.~8, no.~4, pp. 265--274, 2014.
	
	\bibitem{6819451}
	Q.~{Zhang}, Y.~{Tian}, Y.~{Yang}, and C.~{Pan}, ``Automatic spatial–spectral
	feature selection for hyperspectral image via discriminative sparse
	multimodal learning,'' \emph{IEEE Transactions on Geoscience and Remote
		Sensing}, vol.~53, no.~1, pp. 261--279, 2015.
	
	\bibitem{5570985}
	Y.~{Tarabalka}, J.~A. {Benediktsson}, J.~{Chanussot}, and J.~C. {Tilton},
	``Multiple spectral–spatial classification approach for hyperspectral
	data,'' \emph{IEEE Transactions on Geoscience and Remote Sensing}, vol.~48,
	no.~11, pp. 4122--4132, 2010.
	
	\bibitem{1717732}
	Y.~{Bazi} and F.~{Melgani}, ``Toward an optimal svm classification system for
	hyperspectral remote sensing images,'' \emph{IEEE Transactions on Geoscience
		and Remote Sensing}, vol.~44, no.~11, pp. 3374--3385, 2006.
	
	\bibitem{7887259}
	G.~{Cui}, N.~{Li}, L.~{Pallotta}, G.~{Foglia}, and L.~{Kong}, ``Geometric
	barycenters for covariance estimation in compound-gaussian clutter,''
	\emph{IET Radar, Sonar Navigation}, vol.~11, no.~3, pp. 404--409, 2017.
	
	\bibitem{8401913}
	S.~{Han}, A.~{De Maio}, V.~{Carotenuto}, L.~{Pallotta}, and X.~{Huang},
	``Censoring outliers in radar data: An approximate ml approach and its
	analysis,'' \emph{IEEE Transactions on Aerospace and Electronic Systems},
	vol.~55, no.~2, pp. 534--546, 2019.
	
	\bibitem{Huber1964Robust}
	Huber and J.~Peter, ``Robust estimation of a location parameter,'' \emph{Annals
		of Mathematical Statistics}, vol.~35, no.~1, pp. 73--101, 1964.
	
	\bibitem{doi:10.1080/01621459.1974.10482962}
	F.~R. Hampel, ``The influence curve and its role in robust estimation,''
	\emph{Journal of the American Statistical Association}, vol.~69, no. 346, pp.
	383--393, 1974.
	
	\bibitem{10.2307/2241079}
	D.~E. Tyler, ``A distribution-free {M}-estimator of multivariate scatter,''
	\emph{The Annals of Statistics}, vol.~15, no.~1, pp. 234--251, 1987.
	
	\bibitem{10.2307/2957994}
	R.~A. Maronna, ``Robust {M}-estimators of multivariate location and scatter,''
	\emph{The Annals of Statistics}, vol.~4, no.~1, pp. 51--67, 1976.
	
	\bibitem{4359541}
	F.~{Pascal}, Y.~{Chitour}, J.~{Ovarlez}, P.~{Forster}, and P.~{Larzabal},
	``Covariance structure maximum-likelihood estimates in compound {G}aussian
	noise: {E}xistence and algorithm analysis,'' \emph{IEEE Transactions on
		Signal Processing}, vol.~56, no.~1, pp. 34--48, Jan 2008.
	
	\bibitem{6512056}
	M.~{Mahot}, F.~{Pascal}, P.~{Forster}, and J.~{Ovarlez}, ``Asymptotic
	properties of robust complex covariance matrix estimates,'' \emph{IEEE
		Transactions on Signal Processing}, vol.~61, no.~13, pp. 3348--3356, 2013.
	
	\bibitem{6636083}
	M.~{Greco} and F.~{Gini}, ``Cramér-{R}ao lower bounds on covariance matrix
	estimation for complex elliptically symmetric distributions,'' \emph{IEEE
		Transactions on Signal Processing}, vol.~61, no.~24, pp. 6401--6409, 2013.
	
	\bibitem{506799}
	J.~P. {Hoffbeck} and D.~A. {Landgrebe}, ``Covariance matrix estimation and
	classification with limited training data,'' \emph{IEEE Transactions on
		Pattern Analysis and Machine Intelligence}, vol.~18, no.~7, pp. 763--767,
	July 1996.
	
	\bibitem{LEDOIT2004365}
	O.~Ledoit and M.~Wolf, ``A well-conditioned estimator for large-dimensional
	covariance matrices,'' \emph{Journal of Multivariate Analysis}, vol.~88,
	no.~2, pp. 365--411, 2004.
	
	\bibitem{4490277}
	P.~{Stoica}, J.~{Li}, X.~{Zhu}, and J.~R. {Guerci}, ``On using a priori
	knowledge in {S}pace-{T}ime {A}daptive {P}rocessing,'' \emph{IEEE
		Transactions on Signal Processing}, vol.~56, no.~6, pp. 2598--2602, June
	2008.
	
	\bibitem{5484583}
	Y.~{Chen}, A.~{Wiesel}, Y.~C. {Eldar}, and A.~O. {Hero}, ``Shrinkage algorithms
	for {MMSE} covariance estimation,'' \emph{IEEE Transactions on Signal
		Processing}, vol.~58, no.~10, pp. 5016--5029, Oct 2010.
	
	\bibitem{5743027}
	Y.~{Chen}, A.~{Wiesel}, and A.~O. {Hero}, ``Robust shrinkage estimation of
	high-dimensional covariance matrices,'' \emph{IEEE Transactions on Signal
		Processing}, vol.~59, no.~9, pp. 4097--4107, Sep. 2011.
	
	\bibitem{6894189}
	F.~Pascal, Y.~Chitour, and Y.~Quek, ``Generalized robust shrinkage estimator
	and its application to {STAP} detection problem,'' \emph{IEEE Transactions on
		Signal Processing}, vol.~62, no.~21, pp. 5640--5651, Nov 2014.
	
	\bibitem{6879466}
	Y.~Sun, P.~Babu, and D.~P. Palomar, ``Regularized {T}yler's scatter estimator:
	{E}xistence, uniqueness, and algorithms,'' \emph{IEEE Transactions on Signal
		Processing}, vol.~62, no.~19, pp. 5143--5156, Oct 2014.
	
	\bibitem{6913007}
	E.~Ollila and D.~E. Tyler, ``{R}egularized ${M}$-{E}stimators of scatter
	matrix,'' \emph{IEEE Transactions on Signal Processing}, vol.~62, no.~22, pp.
	6059--6070, Nov 2014.
	
	\bibitem{xie2021cross}
	L.~Xie, Z.~He, J.~Tong, J.~Li, and J.~Xi, ``Cross-validated tuning of shrinkage
	factors for mvdr beamforming based on regularized covariance matrix
	estimation,'' \emph{arXiv preprint arXiv:2104.01909}, 2021.
	
	\bibitem{arlot2010survey}
	S.~Arlot, A.~Celisse \emph{et~al.}, ``A survey of cross-validation procedures
	for model selection,'' \emph{Statistics surveys}, vol.~4, pp. 40--79, 2010.
	
	\bibitem{TONG2018223}
	J.~Tong, R.~Hu, J.~Xi, Z.~Xiao, Q.~Guo, and Y.~Yu, ``Linear shrinkage
	estimation of covariance matrices using low-complexity cross-validation,''
	\emph{Signal Processing}, vol. 148, pp. 223--233, 2018.
	
	\bibitem{7422135}
	J.~{Tong}, P.~J. {Schreier}, Q.~{Guo}, S.~{Tong}, J.~{Xi}, and Y.~{Yu},
	``Shrinkage of covariance matrices for linear signal estimation using
	cross-validation,'' \emph{IEEE Transactions on Signal Processing}, vol.~64,
	no.~11, pp. 2965--2975, June 2016.
	
	\bibitem{1310021}
	G.~Alfano, A.~D. Maio, and E.~Conte, ``Polarization diversity detection of
	distributed targets in compound-{G}aussian clutter,'' \emph{IEEE Transactions
		on Aerospace and Electronic Systems}, vol.~40, no.~2, pp. 755--765, April
	2004.
	
	\bibitem{7299689}
	J.~{Liu}, W.~{Liu}, B.~{Chen}, H.~{Liu}, H.~{Li}, and C.~{Hao}, ``Modified
	{R}ao test for multichannel adaptive signal detection,'' \emph{IEEE
		Transactions on Signal Processing}, vol.~64, no.~3, pp. 714--725, Feb 2016.
	
	\bibitem{CUI2012430}
	G.~Cui, L.~Kong, X.~Yang, and J.~Yang, ``Distributed target detection with
	polarimetric {MIMO} radar in compound-{G}aussian clutter,'' \emph{Digital
		Signal Processing}, vol.~22, no.~3, pp. 430--438, 2012.
	
	\bibitem{6298979}
	A.~{Wiesel}, ``Geodesic convexity and covariance estimation,'' \emph{IEEE
		Transactions on Signal Processing}, vol.~60, no.~12, pp. 6182--6189, 2012.
	
	\bibitem{7439863}
	Y.~{Sun}, P.~{Babu}, and D.~P. {Palomar}, ``Robust estimation of structured
	covariance matrix for heavy-tailed elliptical distributions,'' \emph{IEEE
		Transactions on Signal Processing}, vol.~64, no.~14, pp. 3576--3590, July
	2016.
	
	\bibitem{8450037}
	A.~{De Maio}, L.~{Pallotta}, J.~{Li}, and P.~{Stoica}, ``Loading factor
	estimation under affine constraints on the covariance eigenvalues with
	application to radar target detection,'' \emph{IEEE Transactions on Aerospace
		and Electronic Systems}, vol.~55, no.~3, pp. 1269--1283, 2019.
	
	\bibitem{6557488}
	Y.~I. {Abramovich} and O.~{Besson}, ``Regularized covariance matrix estimation
	in complex elliptically symmetric distributions using the expected likelihood
	approach— part 1: The over-sampled case,'' \emph{IEEE Transactions on
		Signal Processing}, vol.~61, no.~23, pp. 5807--5818, 2013.
	
	\bibitem{9054969}
	X.~{Du}, A.~{Aubry}, A.~{De Maio}, and G.~{Cui}, ``Toeplitz structured
	covariance matrix estimation for radar applications,'' \emph{IEEE Signal
		Processing Letters}, vol.~27, pp. 595--599, 2020.
	
	\bibitem{8747386}
	J.~{Li}, A.~{Aubry}, A.~{De Maio}, and J.~{Zhou}, ``An el approach for
	similarity parameter selection in ka covariance matrix estimation,''
	\emph{IEEE Signal Processing Letters}, vol.~26, no.~8, pp. 1217--1221, 2019.
	
	\bibitem{6867011}
	A.~{Aubry}, V.~{Carotenuto}, A.~D. {Maio}, and G.~{Foglia}, ``Exploiting
	multiple a priori spectral models for adaptive radar detection,'' \emph{IET
		Radar, Sonar Navigation}, vol.~8, no.~7, pp. 695--707, 2014.
	
	\bibitem{892662}
	M.~{Steiner} and K.~{Gerlach}, ``Fast converging adaptive processor or a
	structured covariance matrix,'' \emph{IEEE Transactions on Aerospace and
		Electronic Systems}, vol.~36, no.~4, pp. 1115--1126, Oct 2000.
	
	\bibitem{6558039}
	A.~{Aubry}, A.~{De Maio}, and V.~{Carotenuto}, ``Optimality claims for the fml
	covariance estimator with respect to two matrix norms,'' \emph{IEEE
		Transactions on Aerospace and Electronic Systems}, vol.~49, no.~3, pp.
	2055--2057, 2013.
	
	\bibitem{LU2005449}
	N.~Lu and D.~L. Zimmerman, ``The likelihood ratio test for a separable
	covariance matrix,'' \emph{Statistics Probability Letters}, vol.~73, no.~4,
	pp. 449--457, 2005.
	
	\bibitem{7950961}
	Y.~Wang, W.~Xia, Z.~He, H.~Li, and A.~P. Petropulu, ``Polarimetric detection in
	compound gaussian clutter with {K}ronecker structured covariance matrix,''
	\emph{IEEE Transactions on Signal Processing}, vol.~65, no.~17, pp.
	4562--4576, Sept 2017.
	
	\bibitem{823928}
	A.~B. {Kostinski} and A.~C. {Koivunen}, ``On the condition number of gaussian
	sample-covariance matrices,'' \emph{IEEE Transactions on Geoscience and
		Remote Sensing}, vol.~38, no.~1, pp. 329--332, 2000.
	
	\bibitem{7547360}
	Y.~Sun, P.~Babu, and D.~P. Palomar, ``{M}ajorization-minimization algorithms in
	signal processing, communications, and machine learning,'' \emph{IEEE
		Transactions on Signal Processing}, vol.~65, no.~3, pp. 794--816, Feb 2017.
	
	\bibitem{wiesel2015structured}
	A.~Wiesel, T.~Zhang \emph{et~al.}, ``Structured robust covariance estimation,''
	\emph{Foundations and Trends in Signal Processing}, vol.~8, no.~3, pp.
	127--216, 2015.
	
	\bibitem{4454222}
	M.~Hurtado and A.~Nehorai, ``Polarimetric detection of targets in heavy
	inhomogeneous clutter,'' \emph{IEEE Transactions on Signal Processing},
	vol.~56, no.~4, pp. 1349--1361, April 2008.
	
	\bibitem{1561887}
	A.~{De Maio}, ``Robust adaptive radar detection in the presence of steering
	vector mismatches,'' \emph{IEEE Transactions on Aerospace and Electronic
		Systems}, vol.~41, no.~4, pp. 1322--1337, Oct 2005.
	
	\bibitem{249129}
	L.~M. Novak, M.~C. Burl, and W.~W. Irving, ``Optimal polarimetric processing
	for enhanced target detection,'' \emph{IEEE Transactions on Aerospace and
		Electronic Systems}, vol.~29, no.~1, pp. 234--244, Jan 1993.
	
	\bibitem{622504}
	L.~C. {Godara}, ``Application of antenna arrays to mobile communications. {II}.
	{B}eam-forming and direction-of-arrival considerations,'' \emph{Proceedings
		of the IEEE}, vol.~85, no.~8, pp. 1195--1245, 1997.
	
	\bibitem{habets2009new}
	E.~A.~P. Habets, J.~Benesty, I.~Cohen, S.~Gannot, and J.~Dmochowski, ``New
	insights into the mvdr beamformer in room acoustics,'' \emph{IEEE
		Transactions on Audio, Speech, and Language Processing}, vol.~18, no.~1, pp.
	158--170, 2009.
	
	\bibitem{479429}
	J.~Ward, ``Space-time adaptive processing for airborne radar,'' in \emph{1995
		International Conference on Acoustics, Speech, and Signal Processing},
	vol.~5, May 1995, pp. 2809--2812 vol.5.
	
	\bibitem{book2006}
	R.~Klemm, \emph{{P}rinciples of Space-Time Adaptive Processing}, 01 2006.
	
	\bibitem{1263229}
	W.~L. Melvin, ``{A} {STAP} overview,'' \emph{IEEE Aerospace and Electronic
		Systems Magazine}, vol.~19, no.~1, pp. 19--35, Jan 2004.
	
	\bibitem{9052470}
	L.~{Xie}, Z.~{He}, J.~{Tong}, and W.~{Zhang}, ``A recursive angle-doppler
	channel selection method for reduced-dimension space-time adaptive
	processing,'' \emph{IEEE Transactions on Aerospace and Electronic Systems},
	vol.~56, no.~5, pp. 3985--4000, Oct 2020.
	
	\bibitem{9444455}
	J.~Shi, L.~Xie, Z.~Cheng, Z.~He, and W.~Zhang, ``Angle-doppler channel
	selection method for reduced-dimension stap based on sequential convex
	programming,'' \emph{IEEE Communications Letters}, pp. 1--1, 2021.
	
	\bibitem{davis2007information}
	J.~V. Davis, B.~Kulis, P.~Jain, S.~Sra, and I.~S. Dhillon,
	``Information-theoretic metric learning,'' in \emph{Proceedings of the 24th
		international conference on Machine learning}, 2007, pp. 209--216.
	
	\bibitem{dhillon2008log}
	I.~S. Dhillon, ``The log-determinant divergence and its applications,'' in
	\emph{Householder Symposium XVII, Zeuthen, Germany}, 2008.
	
	\bibitem{hunter2004tutorial}
	D.~R. Hunter and K.~Lange, ``A tutorial on {MM} algorithms,'' \emph{The
		American Statistician}, vol.~58, no.~1, pp. 30--37, 2004.
	
	\bibitem{IPIX}
	``The mcmaster ipix radar sea clutter database,'' Available:
	\url{http://soma.ece.mcmaster.ca/ipix/} [Online], {A}ccessed Jul. 1, 2001.
	
	\bibitem{7539668}
	A.~Breloy, G.~Ginolhac, F.~Pascal, and P.~Forster, ``Robust covariance matrix
	estimation in heterogeneous low rank context,'' \emph{IEEE Transactions on
		Signal Processing}, vol.~64, no.~22, pp. 5794--5806, Nov 2016.
	
	\bibitem{bibby_1974}
	J.~Bibby, ``Axiomatisations of the average and a further generalisation of
	monotonic sequences,'' \emph{Glasgow Mathematical Journal}, vol.~15, no.~1,
	p. 63–65, 1974.
	
	\bibitem{LEDOIT2003603}
	O.~Ledoit and M.~Wolf, ``Improved estimation of the covariance matrix of stock
	returns with an application to portfolio selection,'' \emph{Journal of
		Empirical Finance}, vol.~10, no.~5, pp. 603 -- 621, 2003.
	
	\bibitem{746779}
	T.~L. {Marzetta} and B.~M. {Hochwald}, ``Capacity of a mobile multiple-antenna
	communication link in {R}ayleigh flat fading,'' \emph{IEEE Transactions on
		Information Theory}, vol.~45, no.~1, pp. 139--157, Jan 1999.
	
	\bibitem{Hiai00asymptoticfreeness}
	F.~Hiai and D.~Petz, ``Asymptotic freeness almost everywhere for random
	matrices,'' \emph{Acta Sci. Math. Szeged}, vol.~66, pp. 801--826, 2000.
	
	\bibitem{tulino2004random}
	A.~M. Tulino, S.~Verd{\'u} \emph{et~al.}, ``Random matrix theory and wireless
	communications,'' \emph{Foundations and Trends in Communications and
		Information Theory}, vol.~1, no.~1, pp. 1--182, 2004.
	
\end{thebibliography}

\end{document}